\documentclass[%
aip,
 jmp,%
 amsmath,amssymb,
 reprint,%
 floatfix,
]{revtex4-1}
\pdfoutput=1
\usepackage{color}
\definecolor{indigo}{RGB}{0,0,120}
\usepackage[colorlinks=true, linkcolor=indigo, citecolor=blue, urlcolor=indigo]{hyperref}

\usepackage{graphicx}%
\usepackage{dcolumn}%
\usepackage{bm}%

\begin{document}

\title[Visco-resistive MHD study of internal kink(m=1) modes]{Visco-resistive MHD study of internal kink(m=1) modes}%
\author{J. Mendonca}
 \email{jervis.mendonca@ipr.res.in}
 \affiliation{Institute for Plasma Research, Bhat, Gandhinagar 382428, India}%
 \affiliation{Homi Bhabha National Institute, Training School Complex, Anushakti Nagar, Mumbai -400085, India}
\author{D. Chandra}%
\affiliation{Institute for Plasma Research, Bhat, Gandhinagar 382428, India%
}%
\affiliation{Homi Bhabha National Institute, Training School Complex, Anushakti Nagar, Mumbai -400085, India}

\author{A. Sen}
\affiliation{%
Institute for Plasma Research, Bhat, Gandhinagar 382428, India%
}%
\affiliation{Homi Bhabha National Institute, Training School Complex, Anushakti Nagar, Mumbai -400085, India}

\author{A.Thyagaraja}
 \affiliation{Astrophysics Group, University of Bristol, Bristol, BS8 1TL, UK}

\begin{abstract}
We have investigated the effect of sheared equilibrium flows on the $m = 1, n = 1$ resistive internal kink mode in the framework of a reduced magnetohydrodynamic model in a periodic cylindrical geometry.  %
Our numerical studies show that there is a significant change of the scaling dependence of the mode growth rate on the Lundquist number in the presence of axial flows compared to the no flow case. Poloidal flows do not influence the scaling. We have further found that viscosity  strongly modifies the effect of flows on the (1,1) mode both in the linear and nonlinear regime. Axial flows increase the linear growth rate for low viscosity values, but they decrease the linear growth rate for higher viscosity values. In the case of poloidal flows  the linear growth rate decreases in all cases. Additionally at higher viscosity, we have found strong symmetry breaking in the behaviour of linear growth rates and in the nonlinear saturation levels of the modes as a function of the helicities of the flows.  For axial, poloidal and most helical flow cases, there is flow induced stabilisation of the nonlinear saturation level in the high viscosity regime and destabilisation in the low viscosity regime.  
\end{abstract}
\maketitle

\section{Introduction} \label{introduction}
The $m=1, n=1$ internal kink instability is of great importance in tokamaks and has beeen extensively studied in the past by several authors, notably\cite{VonGoeler1974, Kadomtsev, rosenblu, Porcelli1996, Monticello1986}. The (1,1) mode arises within the q=1 rational surface (where q is the safety factor), when the q at the axis is smaller than 1. It can trigger sawtooth oscillations which can influence plasma quality and confinement\cite{VonGoeler1974,Kadomtsev}. %
Monticello et. al. \cite{Monticello1986} have given an overview of the research on the (1,1) mode and its importance, particularly in the context of research on the sawtooth oscillations.

It is well known that flows are a common occurence in a tokamak, which can be generated intrinsically\cite{Rice2007} or induced externally e.g. by unbalanced NBI injection \cite{Menard2003,Menard2005}. Experiments on NSTX have shown a significant increase of sawtooth period that is attributed to a fast rotation of the plasma\cite{Menard2003,Menard2005}. Experimental studies on sawteeth phenomena in presence of NBI in JET \cite{Chapman2007a, Nave2006}, MAST \cite{Chapman2006a}, and TEXTOR \cite{Chapman2008c} have further shown that there is an asymmetry in sawtooth period depending on the direction of the NBI. The sawtooth period increases with an increase in co-NBI power, and decreases with an increase in counter-NBI power. Thus, these experiments have shown that flow can have a stabilising or destabilising effect on the kink mode depending on the direction of flow.

However, there still does not exist a full understanding of the effect of flows on the $m=1,n=1$ kink instability. A number of past studies have addressed this question. In one of the earliest such studies carried out in a slab geometry, Ofman et. al.\cite{Morrison12} have shown that small flow shear has a stabilising influence on the $m=1$ resistive mode. Numerical studies by Kleva and Guzdar\cite{Guzdar1234} show that toroidal sheared flow close to the sound speed can completely stabilise the (1,1) mode.  Shumlak et al.\cite{Shumlak1995} have also found a similar stabilising effect due to a sheared axial flow on the (1,1) mode in a cylindrical Z-pinch. On the other hand, Gatto et. al. \cite{gatto} have found sheared axial flows to have a destabilising effect on the $m=1$ mode in a reverse field pinch configuration.  
Naitou et al. \cite{naitou_kobayashi_tokuda_1999} have studied the effect of poloidal flow on the kink mode in kinetic and two fluid regimes, and noted a stabilisation of the kink mode that can possibly be related to sawtooth stabilisation. Studies by Mikhailovskii et. al.\cite{Mikhailovskii2008aa}, Wahlberg et. al.\cite{wahl_bond} and Waelbrock\cite{wael12} show that toroidal and poloidal rotations are a stabilising factor for the internal kink mode. %
Chapman et. al.\cite{Chapman2006a} have explained the asymmetry in sawtooth period in terms of the relative direction of the plasma flow with respect to the diamagnetic drift. They postulated that the toroidal component of the diamagnetic drift adds to the toroidal rotation for co-current flow but it reduces the toroidal rotation for counter current flow. %
Therefore, there are conflicting results in the literature regarding the nature of stabilisation due to flows depending on the parameter regime of the studies. Recent analytic calculations by Brunetti et. al.\cite{Brunetti2017} have found that small flow shear has a destabilising effect on the (1,1) mode, but a large flow shear can stabilise it.

It may be noted that most of the past flow studies have been done in the low viscosity regime. However, viscosity can be high in tokamak operations, particularly due to enhancements from turbulent effects and could therefore significantly influence the effect of flow shear on the internal kink mode. For example, Maget et. al.\cite{Maget12}, Wang et. al.\cite{Wang2015}, Tala et. al.\cite{Tala2011} and Takeda et. al.\cite{Takeda2008} have shown that Magnetic Prandtl number in advanced tokamak scenarios can be as high as 100, and stability results in the high viscosity regime can be significantly different from results of the low viscosity regime.  Chen et. al.\cite{Chen1990a} and Ofman et. al.\cite{Ofman1991} have given detailed analytical calculations as to how viscosity can modify shear flow effects for constant-$\psi$ and nonconstant-$\psi$ for the resistive tearing mode instability. Wang et. al.\cite{Wang2015} have also reported from their simulation studies of (2,1) tearing modes in the presence of flows, which showed a destabilisation at lower viscosity and stabilisation at higher viscosity. They predict that this is due to the distortion of magnetic island structures at higher viscosity as reported by Ren et. al.\cite{Ren1999} and La Haye et. al.\cite{Haye2009}. 

Thus, viscosity is an important contributing factor and can change the nature of the effect of flows significantly. In this paper we have addressed this issue and investigated the stability of the (1,1) mode in the presence of sheared flows over a range of viscosity regimes. We indeed find that the high viscosity results are often very different from the low viscosity results. In our study, we have systematically examined the effects of several kinds of sheared flows on the (1,1) mode, namely axial, poloidal and combinations of both kinds of flows in the linear as well as nonlinear regimes.%
Various non-dimensional parameters are used to characterise our results, such as \textbf{S} (the Lundquist number, which measures resistivity), \textbf{Pr} (Prandtl Number, which measures viscosity), \textbf{M}  (Alfv\'{e}n Mach number) and $\lambda$(a measure of the equilibrium flow shear). These linear and nonlinear studies on the (1,1) mode were obtained using the CUTIE code \cite{Thyagaraja2000}.

Our principal findings are as follows. To begin with, we have done the linear scaling studies of the $m=1,n=1$ mode in the absence of flow. Here, the variation of linear growth rates have been studied for different S and Pr values. The obtained scalings are in agreement with past  analytic theory results in the no flow case\cite{Porcelli1987}. With the application of sheared axial flows, a significant change in the scaling of the growth rates is observed. However, in the presence of poloidal flow, there is no such change in scaling as compared to the no flow case. In our linear studies we have noticed that axial flows destabilise the mode in the low viscosity regime, but it stabilises in the high viscosity regime as compared to the no flow case. On the other hand, poloidal flow always tends to stabilise the linear growth rate. For pure axial and poloidal flows, the results do not change if we change the direction of the flow. This symmetry is broken for helical flows where the time evolution of the modes show a significant dependence on the helicity of the flows even in the linear regime.  In the nonlinear regime, there is mostly a reduction of the nonlinear saturation level of the (1,1) mode for both sheared axial and poloidal flows in the high viscosity regime,  while in the low viscosity regime, the poloidal and axial flows are destabilising in nature. Helical flows  show a strong stabilisation for positive helicity and in most cases, weak stabilisation for negative helicity in the high viscosity regime. In the low viscosity regime, this symmetry breaking of helical flow results gets significantly diminished. 

This paper is organised in the following manner. In section \ref{model}, we have described the reduced magnetohydrodynamic(RMHD) model of plasma in a cylindrical geometry. In section \ref{linear results}, we have studied the (1,1) mode in the linear regime. Here, we have described studies of the growth rate scaling in the absence of flow, and compared our results with analytical results from the literature. Then we have repeated these studies in the presence of flow. We have done these studies both in the low and high viscosity regimes. In section \ref{nonlinear results} we have  studied the (1,1) mode  in the nonlinear regime in the absence of flow as well as in presence of axial, poloidal and helical flows. Section \ref{discussion} provides a brief summary and a discussion of the results.

\section{Model} \label{model}

Our numerical investigations have been carried out in the framework of a reduced MHD model in a periodic cylinder geometry $(\rho,\theta,z)$,($\rho$ being the radial coordinate, $\theta$ being the azimuthal coordinate, and $z$ being the axial coordinate) defined in terms of the minor radius, $a$, and  the major radius, $R_{0}$. Using normalised coordinates, we set $\rho = r/a$, $r$ being the radial distance, namely $0\leqslant\rho\leqslant1;0\leqslant\theta,\zeta\leqslant2\pi;\zeta=z/R_{0}$. The model utilises CGS electrostatic units.
We have Fourier expanded the fields, and split the equations into a set of ``mean" equations consisting of the (0,0) components of the field and a set for the ``fluctuating" components consisting of the remaining terms. The mean equations can alternatively be obtained by averaging the full equations over the flux surface. In the linear runs we have solved the mean equations once, while in the nonlinear runs, the mean equations are co-evolved with the equations for fluctuating quantities. The fluctuation equations are :

\begin{equation}
\label{Eq.1}
\begin{aligned}
& \frac{\partial\tilde{W}}{\partial t}+  \mathrm{\mathbf{v_{0}}}\cdot\nabla\tilde{W}  +v_{A}\nabla_{\parallel}\rho_{s}^{2}\nabla_{\perp}^{2}\bar{\psi} \\ = & v_{A}\rho_{s}\frac{1}{r}\frac{\partial\tilde{\psi}}{\partial\theta}\frac{4\pi\rho_{s}}{cB_{0}}j_{0}^{'}  + 
\frac{v_{th}\rho_{s}}{r}\left\{ \tilde{\psi},\rho_{s}^{2}\nabla_{\perp}^{2}\tilde{\psi}\right\}   +\frac{v_{th}\rho_{s}}{r}\left\{ \tilde{W},\tilde{\phi}\right\} \\ &  +\frac{\rho_{s}^{2}W_{0}^{'}}{r}\frac{\partial\tilde{\phi}}{\partial\theta}  +\nu\nabla_{\perp}^{2}\tilde{W} \\
\end{aligned}
\end{equation}

\begin{equation}
\label{Eq.2}
\frac{\partial\tilde{\psi}}{\partial t}+\mathrm{\mathbf{v_{0}}}\cdot\nabla\tilde{\psi}+v_{A}\nabla_{\parallel}\tilde{\phi}=\frac{v_{th}\rho_{s}}{r}\left\{ \tilde{\psi},\tilde{\phi}\right\} +\frac{c^{2}\eta}{4\pi}\nabla_{\perp}^{2}\tilde{\psi}
\end{equation}

where, \[\tilde{W}=\rho_{s}^{2}\nabla\cdot\left(\frac{n_{0}\left(\rho\right)}{n_{0}\left(0\right)}\nabla_{\perp}\tilde{\phi}\right)\]

Equation[\ref{Eq.1}] is the vorticity equation, where $\tilde{W}$ is the perturbed vorticity.  Equation[\ref{Eq.2}] describes the evolution of the perturbed poloidal flux function $\tilde{\psi}$. The resistivity $\eta$ and viscosity $\nu$ are specified quantities and are held constant during our calculations. Additionally, $\rho_{s}=\frac{v_{th}}{\omega_{ci}}$,  where,  $v_{th}^{2}=\left(T_{0i}+T_{0e}\right)/m_{i}$,  $\omega_{ci}=\left(eB_{0}/m_{i}c\right)$, with $T_{0i},T_{0e}$ being ion and electron temperatures respectively. $m_{i}$ is the ion mass, $e$ is the elemenatary charge. $\Phi_{0}(r), \Psi_{0}(r)$ denote the mean electrostatic and magnetostatic potentials respectively.

Also, we have used fixed boundary conditions, along with a conducting boundary. 

This comes from 
\[\frac{\delta\mathbf{E}}{B_{0}}=-\nabla\tilde{\phi}-\frac{1}{c}\frac{\partial\tilde{\phi}}{\partial t}\mathbf{e_{\zeta}}\] where, $\mathbf{E}$ is the electric field, $\phi$ is the electrostatic potential, and  $\mathbf{B_{0}\simeq} B_{0z}{\bf e}_{\zeta}+B_{0\theta}(\rho){\bf e}_{\theta}$  is the equilibrium field.  The fluctuating electric field, $\delta E$, is related to $\tilde{\phi}$ [this has dimensions of length].
We set, $\epsilon=a/R_{0}$, the inverse aspect ratio,  $v_{0}=V_{0z}\left(\rho\right)\mathbf{e}_{\zeta}+a\rho\Omega\left(\rho\right)\mathbf{e}_{\theta}$. The equilibrium axial and poloidal, sub-Alfv\'{e}nic sheared flows are: $M_{z}=V_{0z}/v_{A}$ is the Axial Mach number; $M_{\theta}=\rho\Omega\left(\rho\right)\tau_{A}$ is the poloidal Mach number; $\tau_{A}=a/v_{A}$ the Alfv\'{e}n time; $\tau_{\eta}=(4\pi a^{2}/c^{2}\eta)$ the resistive time; $\tau_{\nu}=(a^{2}/\nu)$ the viscous time. We will use in the following the \textit{Lundquist Number}, $S=\frac{\tau_{\eta}}{\tau_{A}}$, and the \textit{Prandtl Number}, $Pr=\frac{\tau_{\eta}}{\tau_{\nu}}$. 

The velocity perturbations are non-dimensionalised relative to the Alfven speed, $v_{A}=\frac{B_{0}}{\left(4\pi m_{i}n_{0}\right)^{1/2}}$. The magnetic field perturbations are normalised by the equilibrium axial magnetic field $B_{0z}$. The fluctuations of magnetic field and velocity are incompressible in the $(r- \theta)$ plane.

Together, these equations constitute the V-RMHD(visco-resistive MHD) model and we solve them using the CUTIE (\textbf{CU}lham \textbf{T}ransporter of \textbf{I}ons and \textbf{E}lectrons) code \cite{Thyagaraja2000, Chandra2015}. CUTIE is a nonlinear, global, electromagnetic, quasi-neutral, two fluid initial value code. It has been used earlier for studies of tearing modes, ELMs, L to H transitions, internal transport barriers and other problems \cite{Thyagaraja2000,Thyagaraja2010,Chandra2015,Chandra2017}.

\section{Linear Results} \label{linear results}

In this section we describe the results of our linear studies carried out for a q profile of the following form:
\begin{equation}
\label{Eq.3}
q(\rho)=q_{0}\left(1+\left(\frac{\rho}{\rho_{0}}\right)^{2\Lambda}\right)^{\frac{1}{\Lambda}}
\end{equation}  
with the safety factor, $q(\rho)=\frac{\epsilon \rho B_{0z}}{B_{0\theta}(\rho)}$, $q_{0}=0.9$, $\Lambda=1$,  $\rho_{0}=0.6a$, $a=$ radius of the cylinder. %

\begin{center}
\begin{figure}[!htb]
\centering
\includegraphics[scale=0.35]{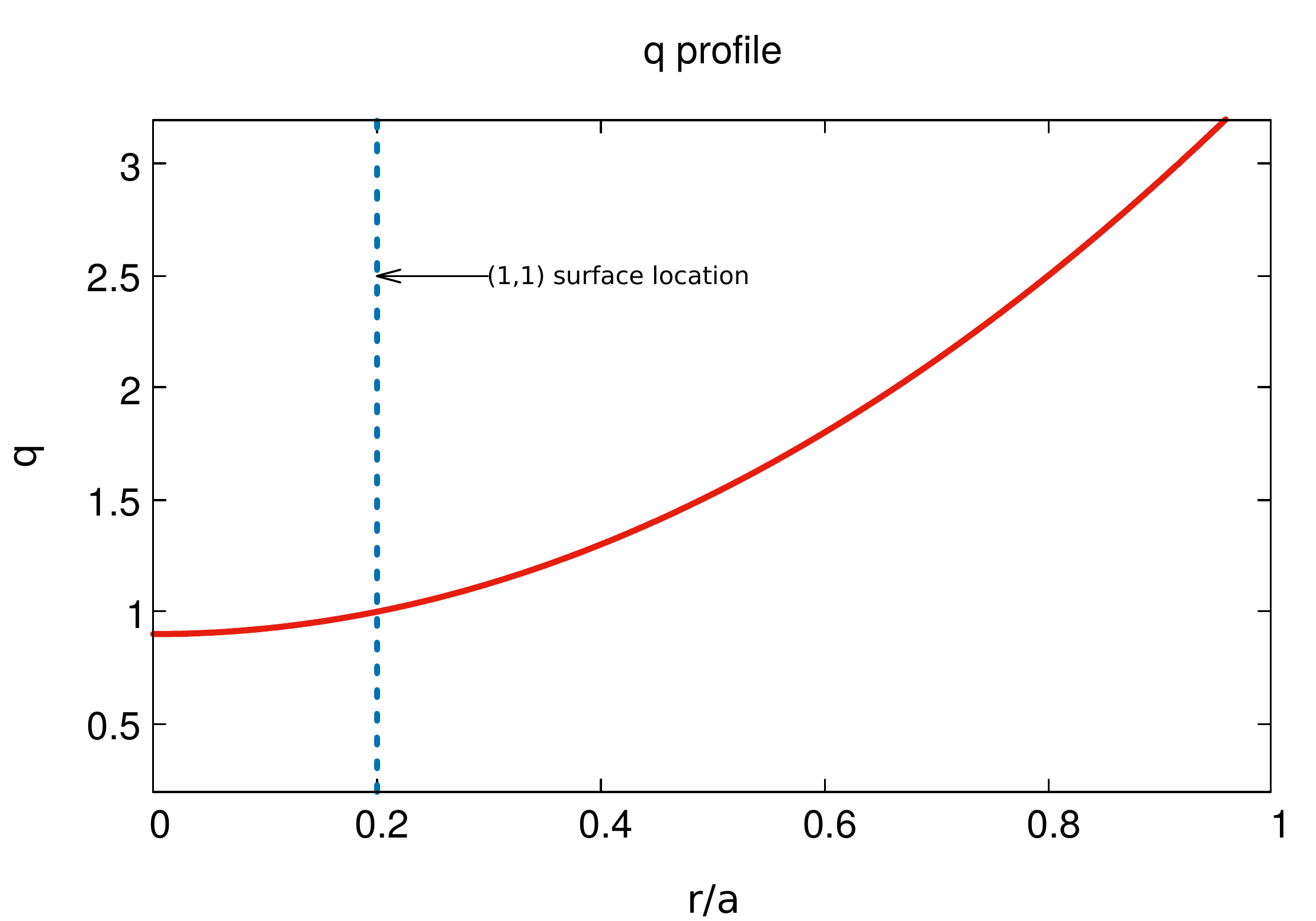}    
\caption{q profile}
\label{fig:q_profile}
\end{figure}
\end{center}

Fig. \ref{fig:q_profile} shows the q profile used in the simulations and the q=1 surface.

\subsection{Scaling with S and Pr}
\begin{center}
\begin{figure}[!htb]
\centering
\includegraphics[scale=0.35]{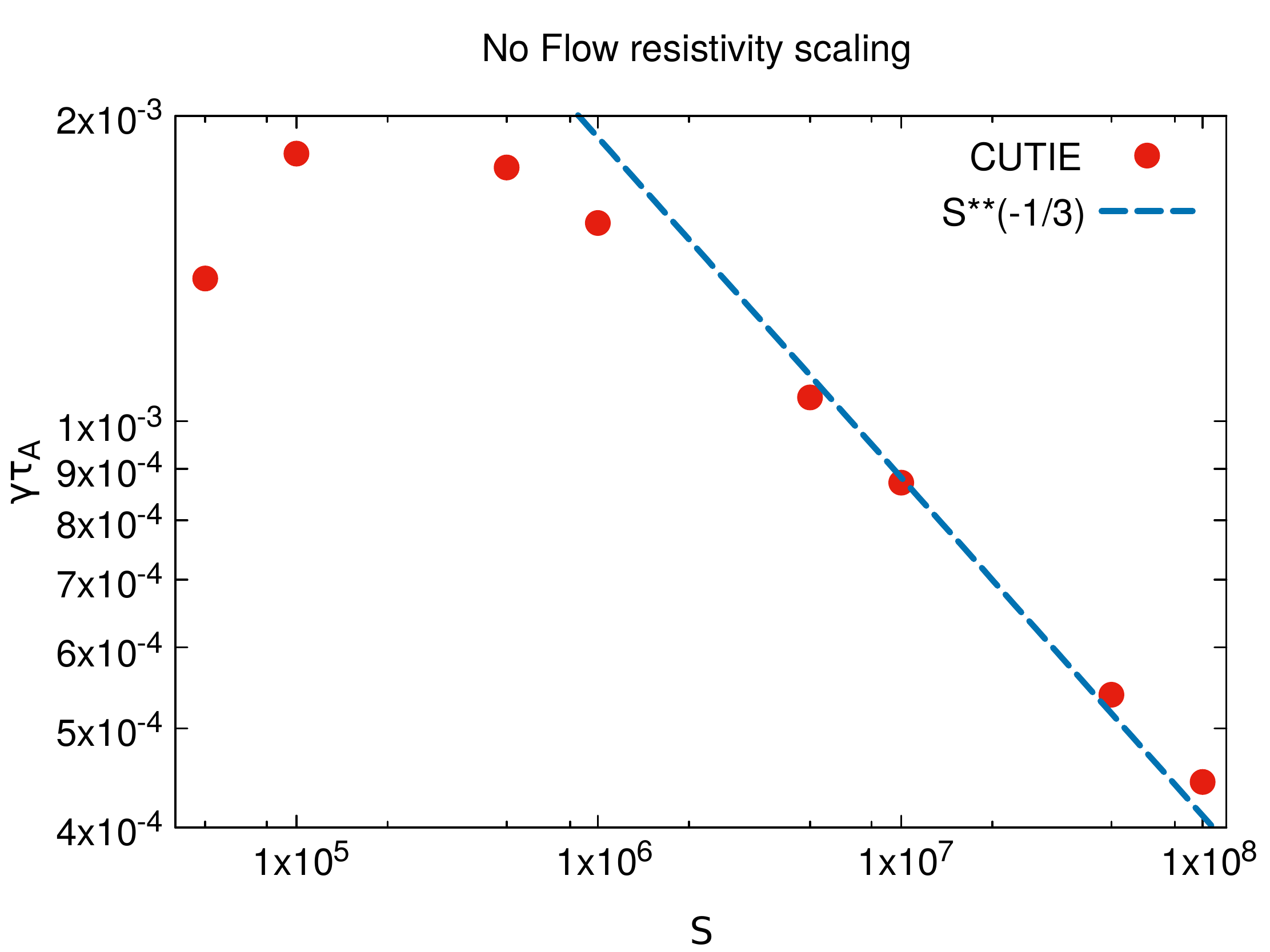}     
\caption{Linear Resistivity scaling without flow for the m=1, n=1 mode at $Pr=0.1$}
\label{fig:res_nf_high}
\end{figure}
\end{center}

At first, we have studied the scaling of the growth rate of the (1,1) mode with S(Lundquist number) and Pr(Prandtl Number). For most of our simulations we have used a flat $\eta$ profile but we get similar results when we use a self-consistent $\eta$ such that $E_{0z}=\eta j_{0z}; E_{0z}=V/(2\pi R_{0})$, where $V$ is the constant loop-voltage. %
In Fig. \ref{fig:res_nf_high},  we have plotted the normalised growth rate $\gamma\tau_{A}$ with $S$, at a fixed $Pr$ of 0.1. We have found a scaling of the variation of $\gamma\tau_{A}$ with $S$ to be of the form $S^{-1/3}$ . These results are  similar to those obtained by Porcelli\cite{Porcelli1987} for the resistive internal kink mode. At low $S$, we notice  a deviation from the scaling that can be attributed to local asymmetries of the equilibrium current density\cite{Militello2004a}. %

\begin{center}
\begin{figure}[!htb]
\centering
\includegraphics[scale=0.35]{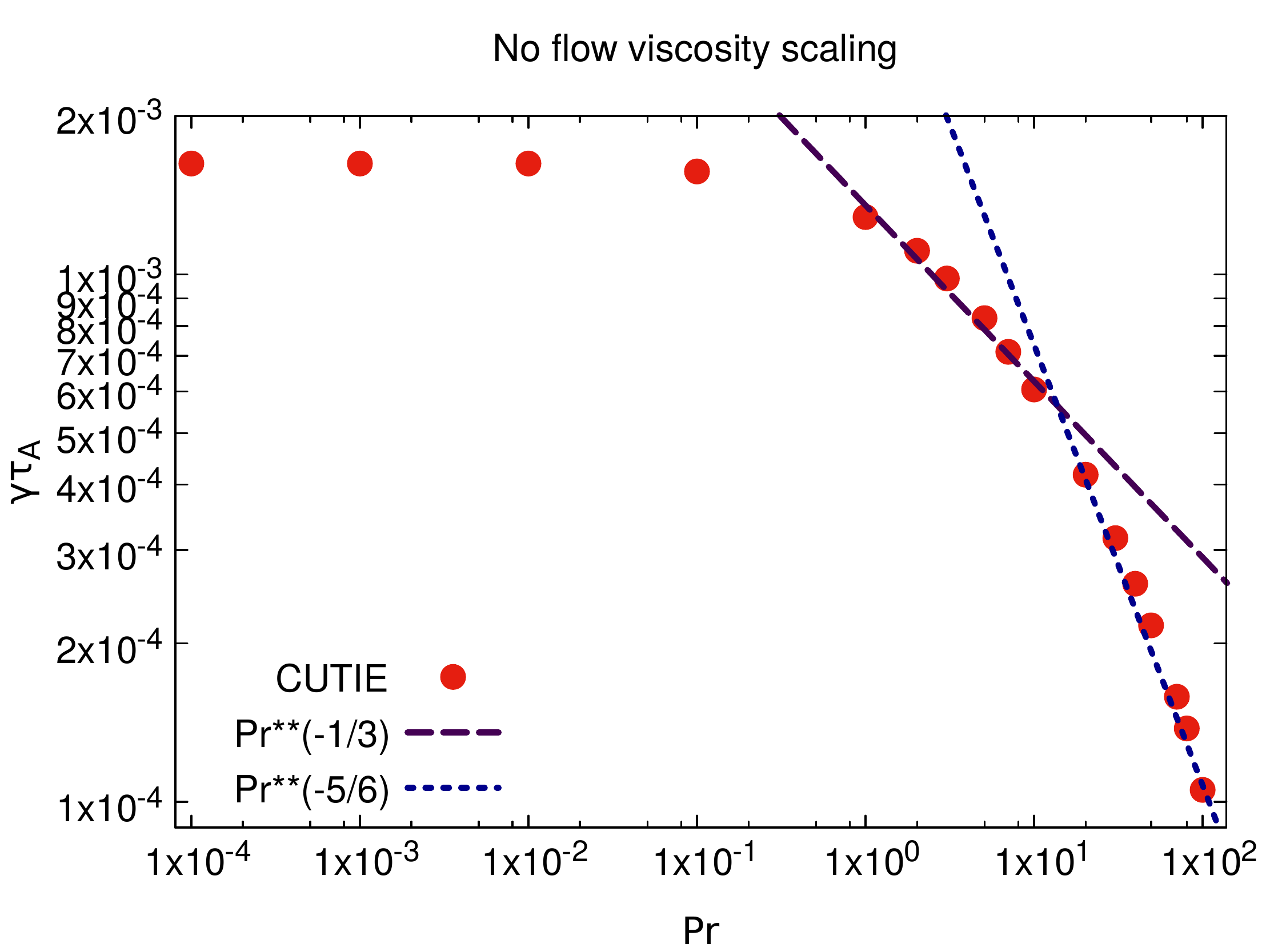} 
\caption{Linear Viscosity scaling without flow for the m=1, n=1 mode at $S=10^6$}
\label{fig:vis_nf_l}
\end{figure}
\end{center}

In Fig. \ref{fig:vis_nf_l}, we observe a $Pr^{-1/3}$ scaling of $\gamma\tau_{A}$ as we vary $Pr$ by keeping $S$ fixed. This scaling agrees with that reported earlier by Porcelli\cite{Porcelli1987}. However, as we increase the viscosity further, the growth rate scaling changes to $Pr^{-5/6}$, which is a new result that has not been reported earlier. It shows that high viscosity can strongly influence the linear growth rate of the modes. 
These results can be qualitatively understood by a standard dominant balance analysis of the dynamical equations of the mode in the inner layer. From the set of model equations (\ref{Eq.1}) and (\ref{Eq.2}) one can obtain the following set of linear inner layer equations:\\

\begin{eqnarray}
(\gamma +iv_{0,res}^{\prime}x)\frac{d^2 \phi}{dx^2} +i\frac{q^{\prime}_{res}}{q_{res}}x \frac{d^2\psi}{dx^2} &=& \nu \frac{d^4 \phi}{dx^4} \label{vort}\\
(\gamma +iv_{0,res}^{\prime}x)\psi + i\frac{q^{\prime}_{res}}{q_{res}}x \phi &=& \eta \frac{d^2 \psi}{dx^2}
\label{Ohm}
\end{eqnarray}
where all quantities are suitably made non-dimensional and where $\nu$, $\eta$ are non-dimensional viscous and resistive diffusivities respectively.  $v_{0,res}^{\prime}$ and $q^{\prime}_{res}$ are the derivatives of the flow terms and q profile respectively at the resonant surface and $\gamma$ is the normalized growth rate. In the absence of flow and in the regime where both the viscous and resistive contributions are important, the term proportional to $\nu$ with the highest derivative dominates over the term proportional to $\gamma$  in (\ref{vort}), while in (\ref{Ohm}) all terms contribute equally. Using the dominant balance argument one then gets,

$$ \gamma \sim \eta^{2/3}\nu^{-1/3} \sim Pr^{-1/3}$$
and the layer width goes as 
$$ x \sim \eta^{1/6}\nu^{1/6} $$ 
This agrees with the numerical scaling obtained in Fig. 3 for moderate values of Pr and is also in accordance with the scaling discussed by Porcelli \cite{Porcelli1987}. For higher values of viscosity, when viscous effects dominate over resistive contributions, the term on the R.H.S. of (\ref{Ohm}) may be ignored in the dominant balance calculation. In this limit the layer width also has a very weak dependence on viscosity and can be nearly taken to be a constant. The balance arguments then lead to $ \gamma \nu \sim x^4$ and hence $\gamma \sim \nu^{-1}$. This scaling is close to the  $\gamma \sim Pr^{-5/6}$ obtained from our numerical solutions. Such a scaling has also been alluded to by Porcelli \cite{Porcelli1987} for the so-called visco-ideal limit. The growth rate becomes nearly constant in the low Pr regime, as we would expect the plasma to be nearly inviscid. We next consider the effect of flows on the linear growth rates.

\subsubsection{Axial Flow}\label{linear axial}

\begin{center}
\begin{figure}[!htb]
\centering
\includegraphics[scale=0.35]{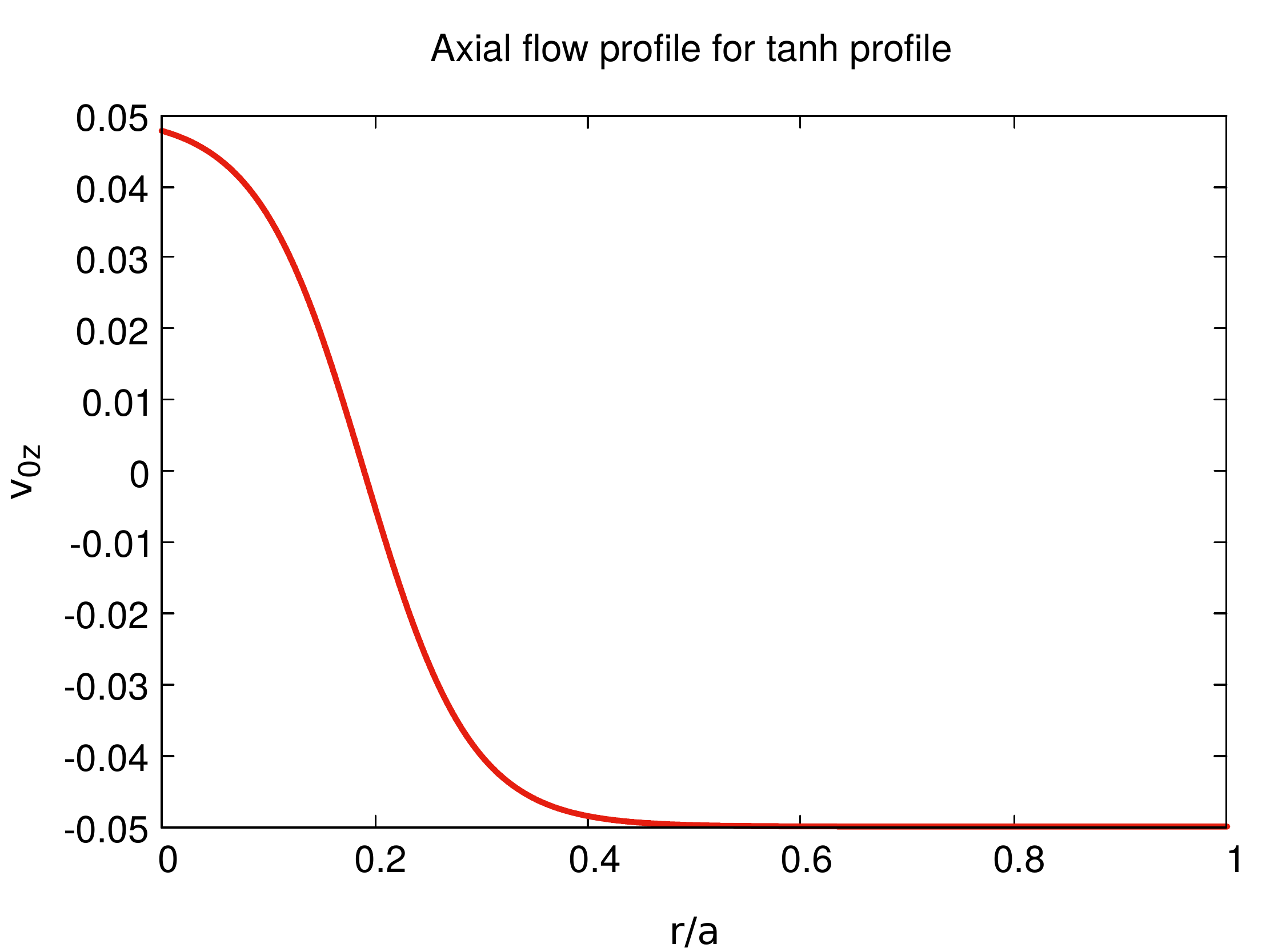}   
\caption{Axial flow profile(tanh profile)}
\label{axial_linear_profile}
\end{figure}
\end{center}

\begin{center}
\begin{figure}[!htb]
\centering
\includegraphics[scale=0.35]{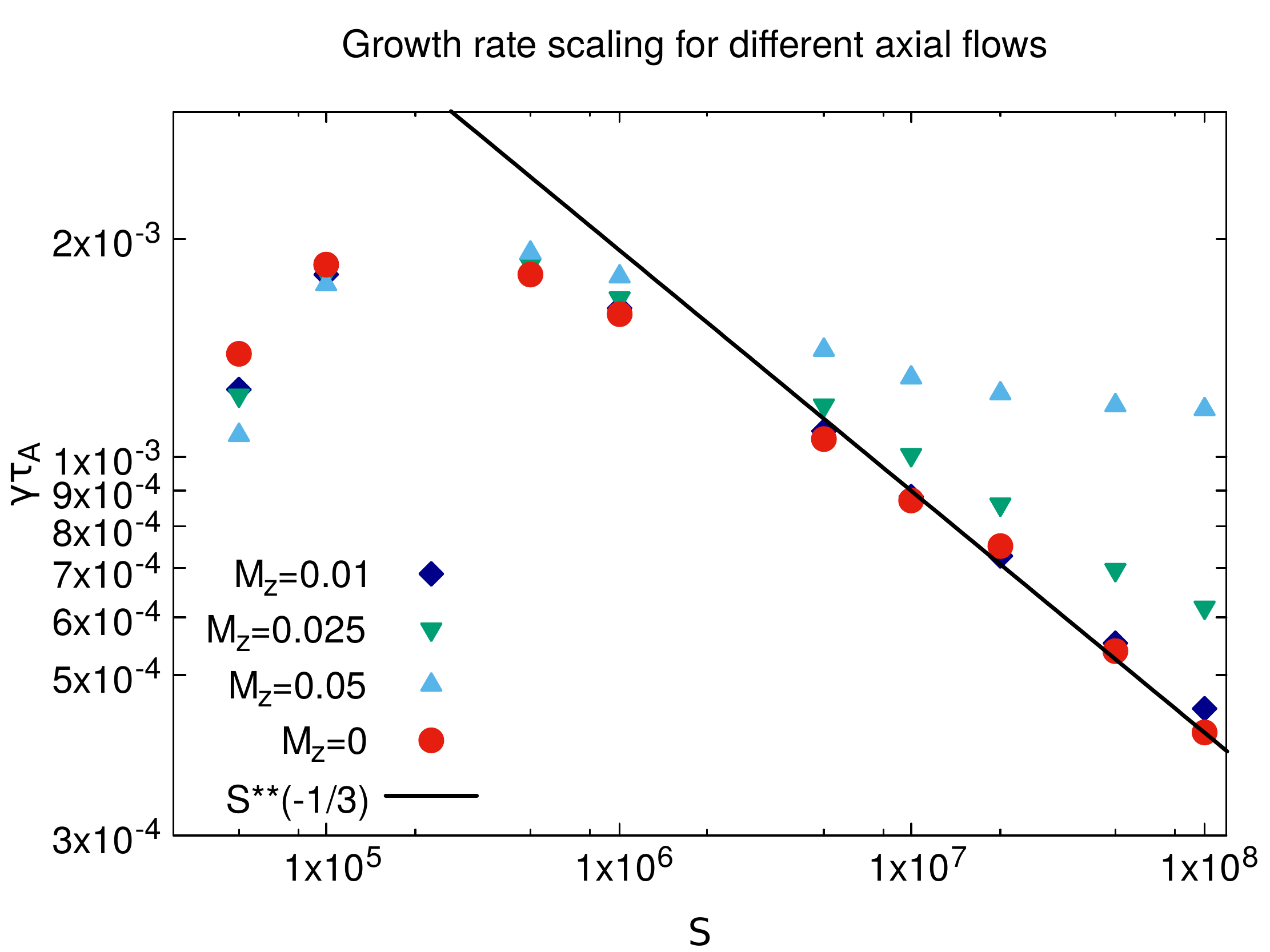} 
\caption{Linear Resistivity scaling with axial flow for the m=1, n=1 mode at $Pr=0.1$ with  $M_{z}=(0.0,0.01,0.025,0.05)$}
\label{res_ax_flow_lin}
\end{figure}
\end{center}

We next present scaling results in the presence of a sheared axial flow. We have used an axial flow profile of the form:

\begin{equation}
\frac{V_{0z}}{v_{A}}=M_{z}\tanh[\lambda(\rho-\rho_{res})]
\end{equation}

where, $V_{0z}$ is the equilibrium axial flow,  $M_{z}$ is the axial Mach number, $\lambda$ is the shear parameter and  $\rho_{res}$ is the location of the mode resonant surface. The flow profile is shown in Fig. \ref{axial_linear_profile}
This profile has the property that it has zero flow and non-zero shear at the resonant surface and hence is a useful profile to study the effect of shear on the mode. It has been used in the past to understand the effect of flow shear on tearing modes\cite{Ofman1991,Chandra2015}. For our linear scaling studies we have taken several different values $M_{z}=0.05$  that are within a physically reasonable range of values for experimental observations. 
In general, the presence of an axial sheared flow has a destabilizing influence on the $m=1$ resistive kink mode,
as has been noted before \cite{Gimblett1996} and is due to the additional ideal free energy arising from the nature of the flow profile. A principal consequence of this is an increase in the growth rate of the kink mode compared to the no flow case. This is clearly seen in Fig. \ref{res_ax_flow_lin} where we have marked the values of the growth rate for the no flow case and for several different finite values of the axial flow (and correspondingly different velocity shears) in a single plot. It is also seen that there is a near independence of the growth rate on $S$ for higher values of $S$. This can be physically understood as follows: as the resistivity decreases (S increases) the growth rate of the classical resistive kink mode decreases 
and the growth is dominated by the ideal driving term of the flow shear. This term is independent of $S$ and hence at higher values of $S$ the growth rate becomes independent of $S$. In Fig.~(\ref{res_ax_flow_lin}) we thus see how an increase in $M_{z}$ progressively changes the $S^{-1/3}$ scaling in the ``no-flow'' case to one independent of $S$.

\begin{center}
\begin{figure}[!htb]
\centering
\includegraphics[scale=0.35]{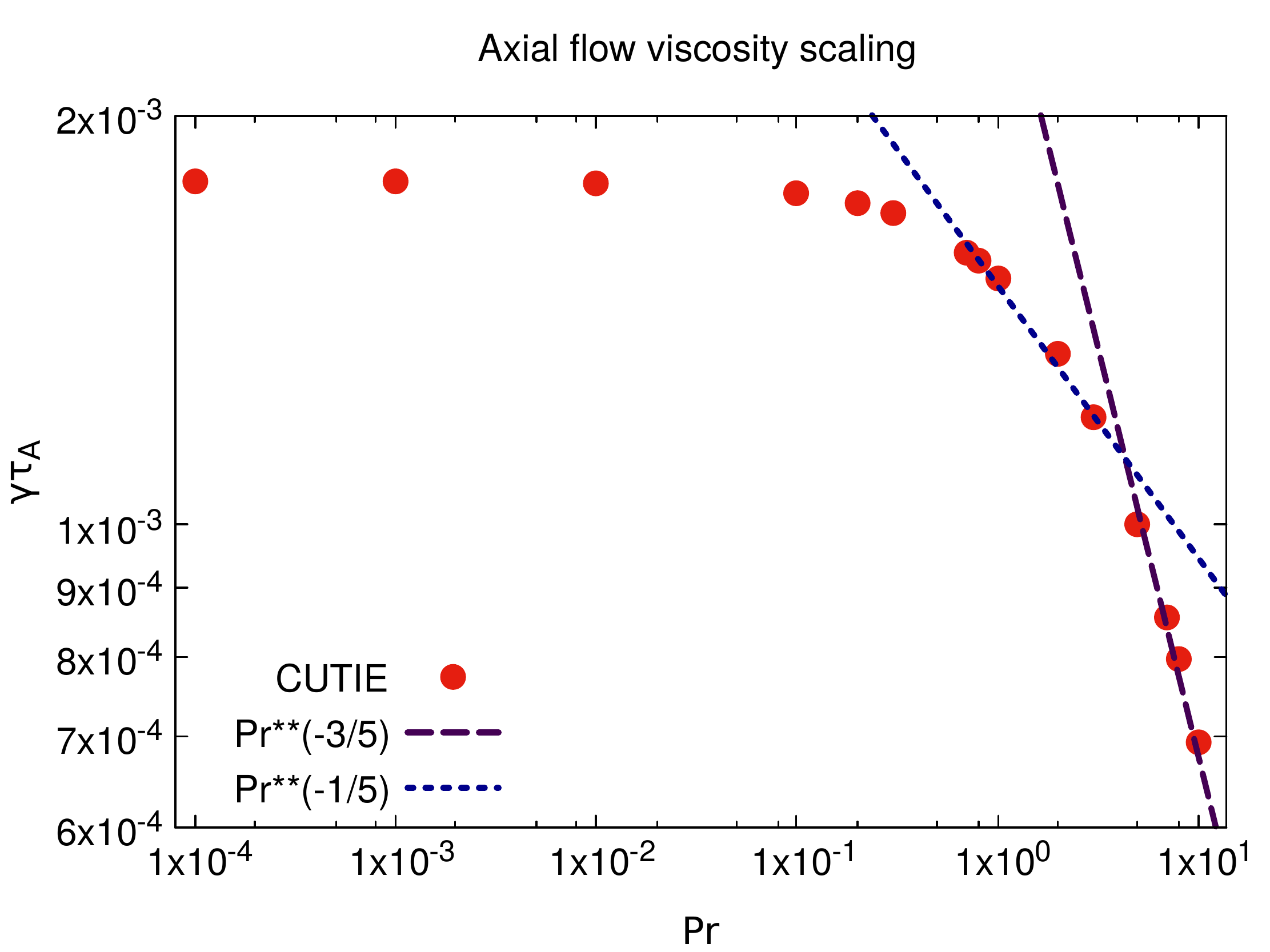} 
\caption{Linear Viscosity scaling with axial flow for the m=1, n=1 mode at $S=10^6$, $M_{z}=0.05$}
\label{vis_ax_flow_lin}
\end{figure}
\end{center}

We have similarly observed a change in the viscosity scaling due to the presence of axial flow and this is shown in Fig. \ref{vis_ax_flow_lin}. As we go from $Pr =1$ to $Pr =10 $, the scaling of the linear growth rate gradually changes from $Pr^{-1/5}$ to $Pr^{-3/5}$ and beyond $Pr =10$, the growth rate becomes negative. If we compare it with the no flow case, there the scaling goes as $Pr^{-1/3}$ up to $Pr=10$, beyond which it changes to $Pr^{-5/6}$. Thus in the presence of an axial flow, the stabilising influence of viscosity is enhanced and can lead to complete stabilisation of the $m=1$ visco-resistive mode at high $Pr$ numbers.

\subsubsection{Poloidal flow} 

For our poloidal flow studies, we have used the following flow profile,
\begin{equation}
\frac{V_{0\theta}}{v_{A}}=M_{\theta}(\rho)
\end{equation}
 
where,
\[M_{\theta}(\rho)=\Omega\tau_{A}\;\rho\left(1+k\rho\right)
 \]
 
Here, $V_{0\theta}$ is the equilibrium poloidal flow, and $\Omega$ is the
poloidal angular frequency and $k$ measures the shear in the flow. 

\begin{center}
\begin{figure}[!htb]
\centering
\includegraphics[scale=0.35]{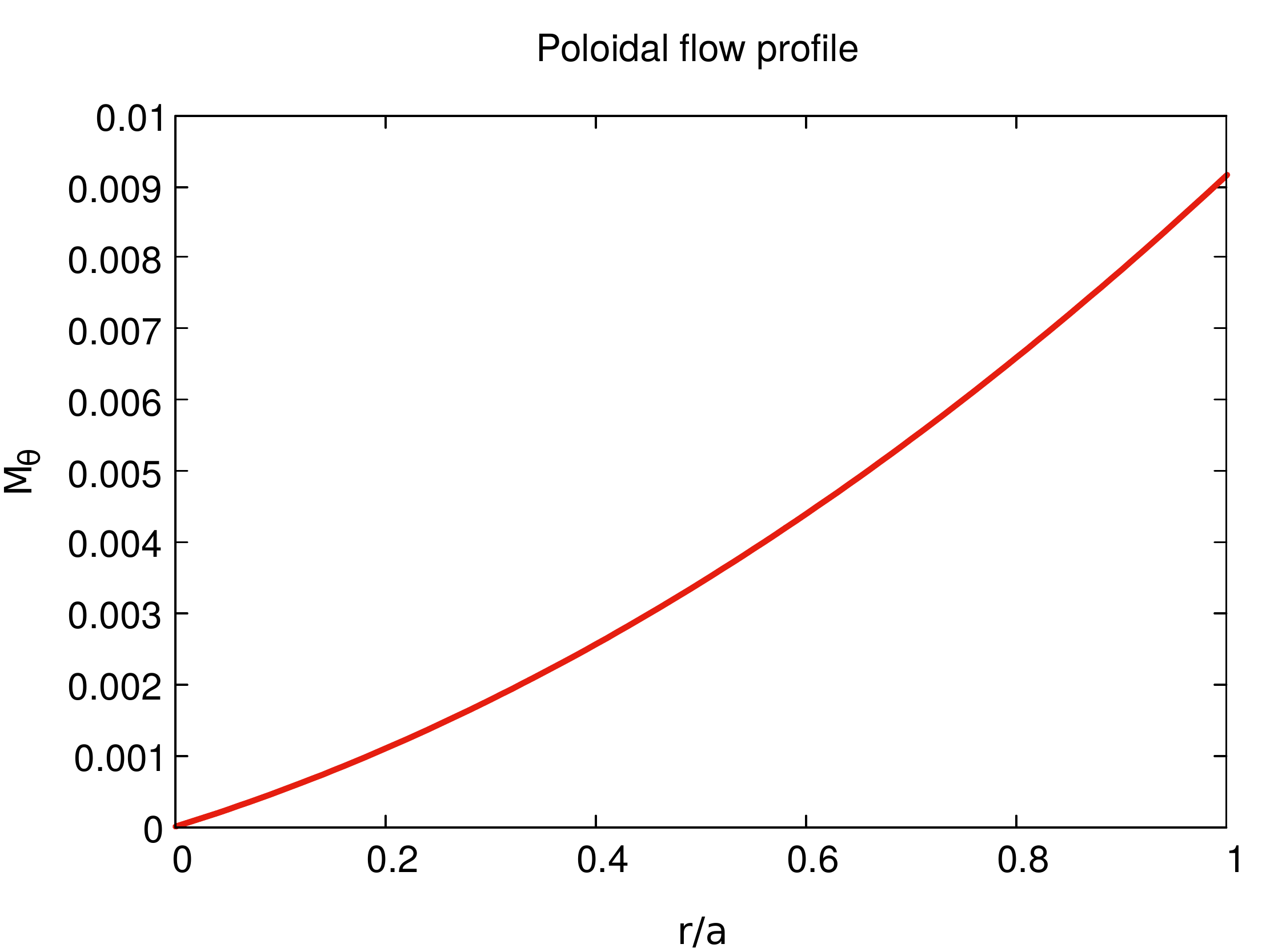}  
\caption{Poloidal flow profile}
\label{poloidal_profile}
\end{figure}
\end{center}

The profile is plotted in Fig. \ref{poloidal_profile}

\begin{center}
\begin{figure}[!htb]
\centering
\includegraphics[scale=0.35]{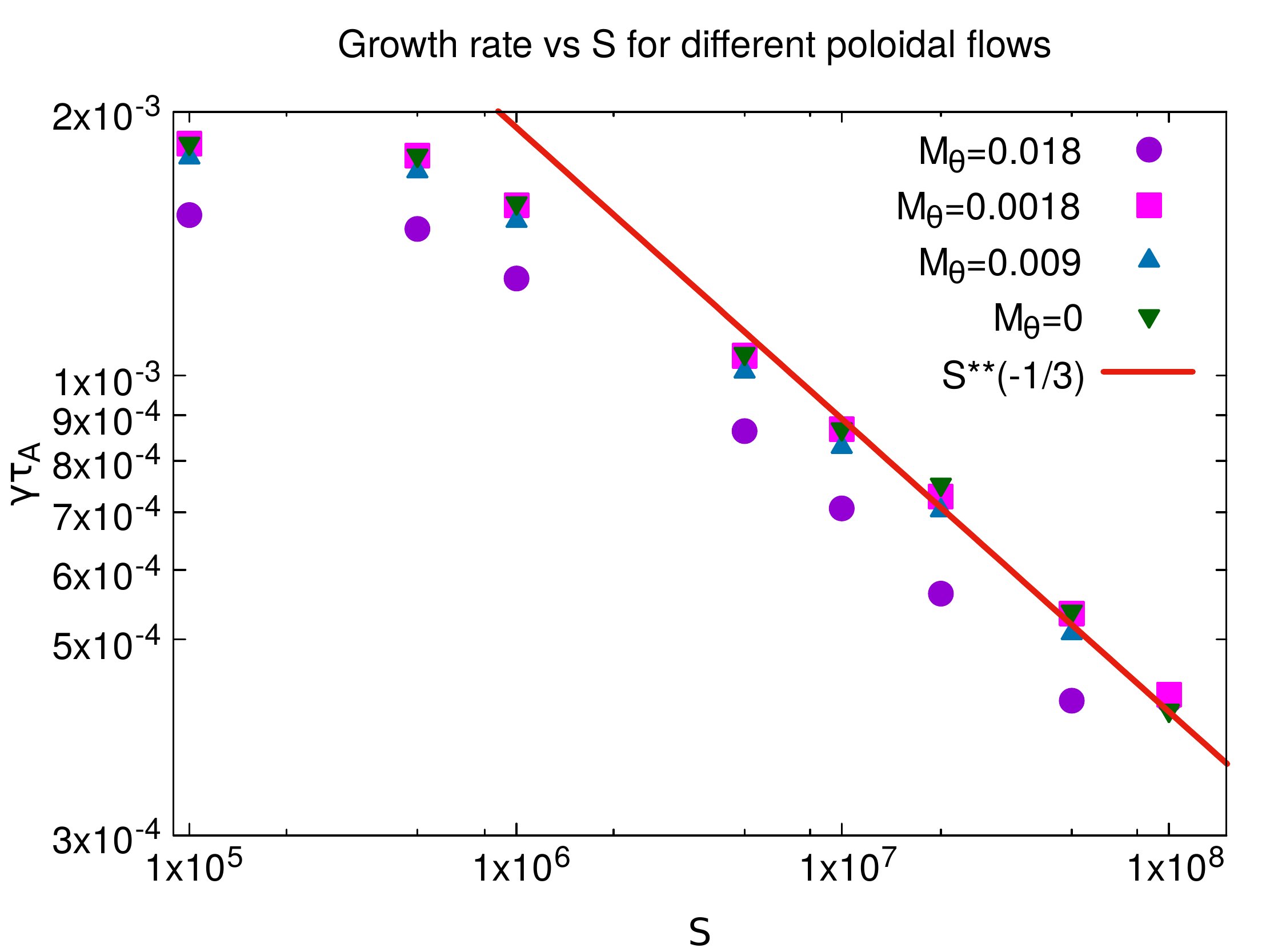}  
\caption{Linear Resistivity scaling with poloidal flow for the m=1,n=1 mode at $Pr=0.1$,  $M_{\theta}=(0.0,0.0018,0.009,0.018)$}
\label{rmhd_pol_resis_scal}
\end{figure}
\end{center}

We have obtained the growth rates for several different values of $M_{\theta}$ and the results are plotted in Fig.~( \ref{rmhd_pol_resis_scal}). In contrast to the sheared axial flow the sheared poloidal flow has a stabilizing influence on the resistive kink mode so that it decreases the value of the growth rate. This stabilizing influence is independent of $S$ and hence the net effect is a shift in the value of the growth rate without a change in the scaling dependence on $S$ which remains the same as the no-flow case, namely $\gamma\tau_{A}\propto S^{-1/3}$. As seen in Fig.~(\ref{rmhd_pol_resis_scal}) as the poloidal Mach number is increased from zero, the scaling curve shifts downwards without changing shape. For very low values of $M_{\theta}$, the curve almost coincides with the  no-flow case, as expected. 
 
\begin{center}
\begin{figure}[!htbp]
\centering
\includegraphics[scale=0.35]{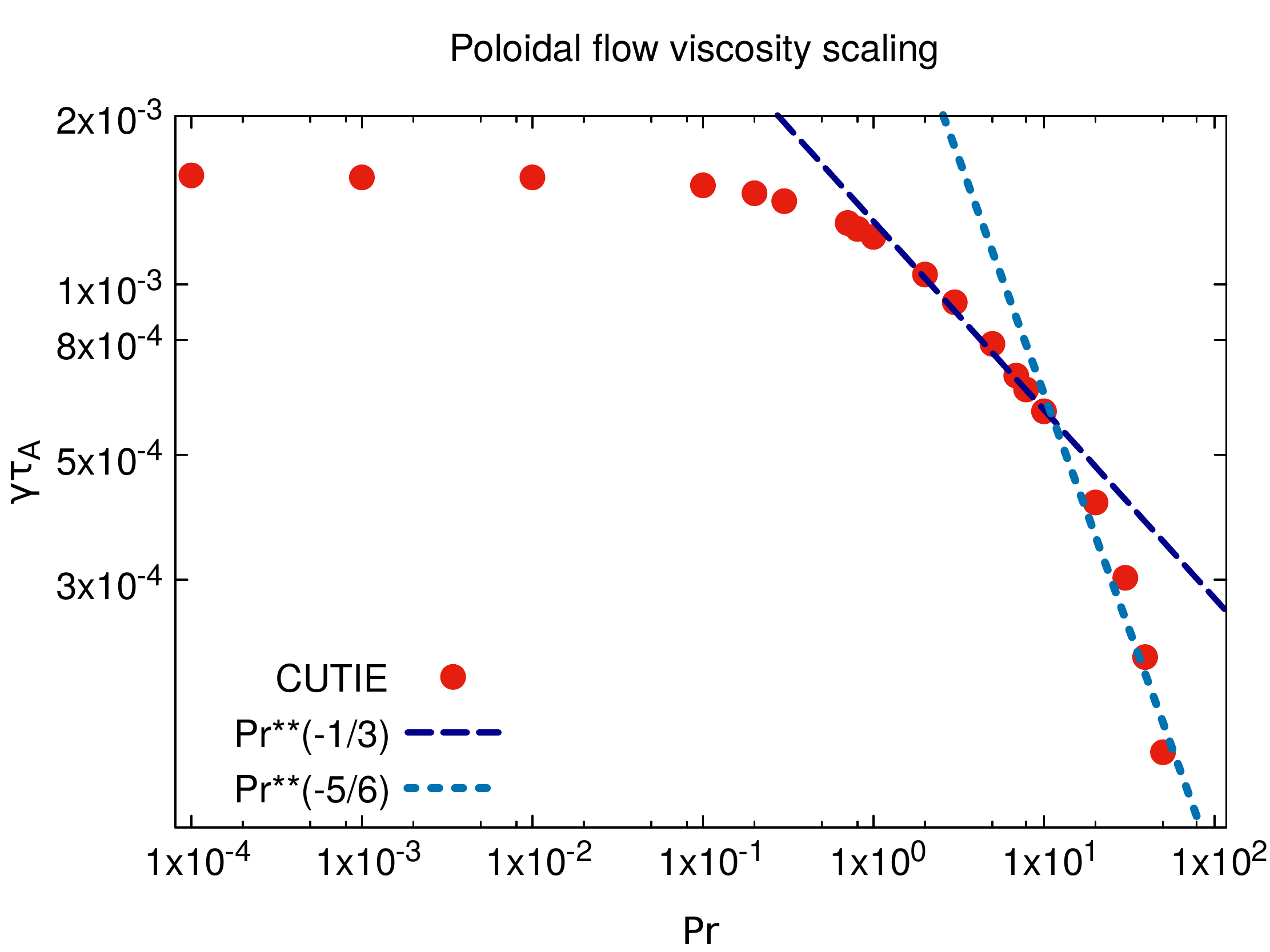}  
\caption{Linear Viscosity scaling with poloidal flow for the m=1,n=1 mode at $S=10^6$,  $M_{\theta}=0.009$}
\label{rmhd_pol_vis_scal}
\end{figure}
\end{center}

In Fig. \ref{rmhd_pol_vis_scal} we have displayed the scaling of the growth rate with viscosity for a fixed value of the resistivity. We find that the scaling is similar to the no flow case, namely $\gamma\tau_{A} \propto Pr^{-1/3}$ at a lower viscosity and  $\gamma\tau_{A} \propto Pr^{-5/6}$  at a higher viscosity. 

\subsection{Effect of flows in different viscosity regimes}

\begin{figure}[!htb]
\centering
\includegraphics[scale=0.35]{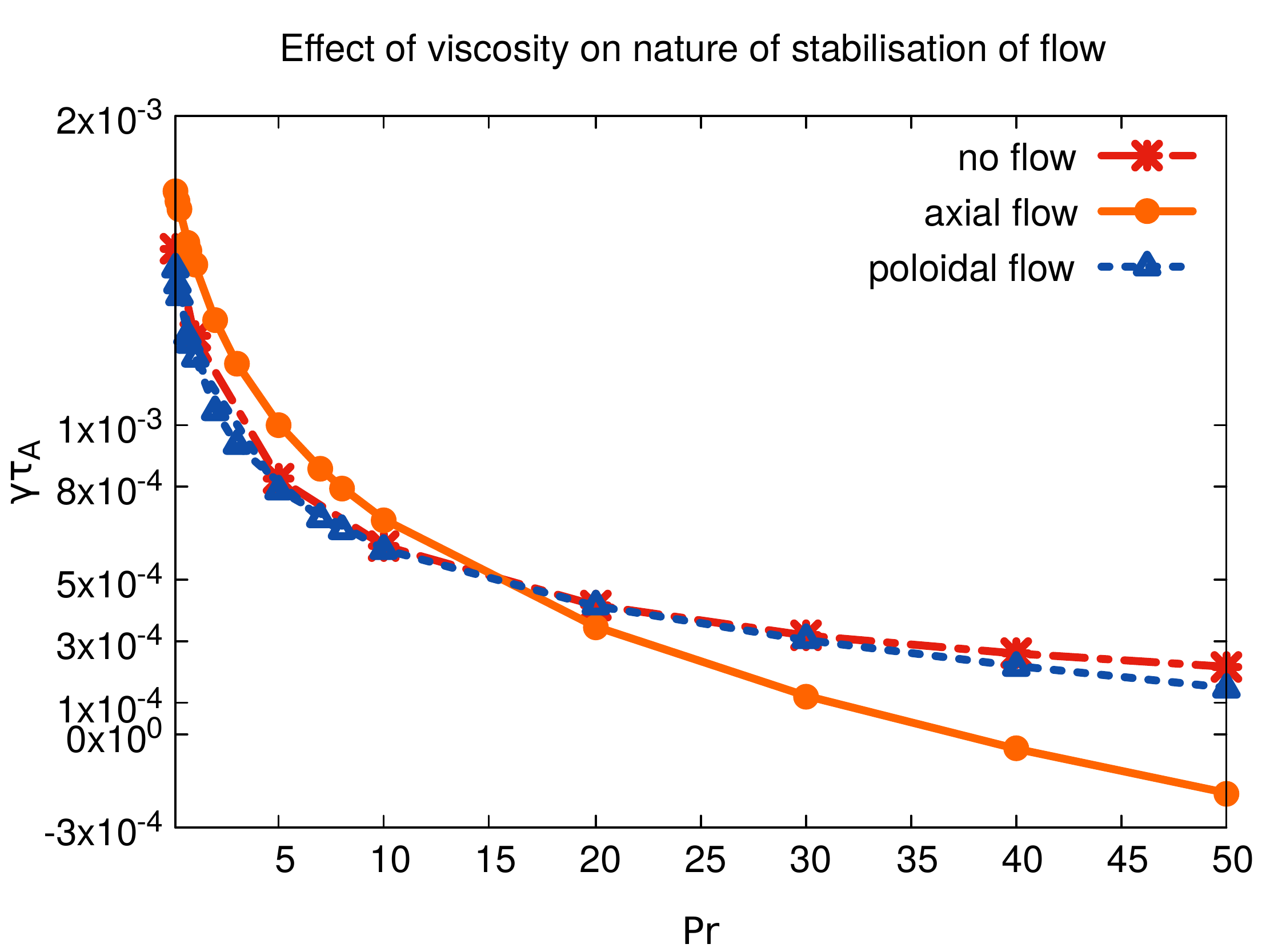}  
\caption{Effect of viscosity on the linear growth rate of the m=1, n=1 mode with and without axial flow at $S=10^6$}
\label{high_vis_flow_comparison}
\end{figure}

In Fig. \ref{high_vis_flow_comparison}, we compare the linear growth rate changes of the $m=1, n=1$ mode with axial and poloidal flows as we go from low to high viscosity.  We note that the nature of stabilisation for axial flows changes as we increase the viscosity. While axial flows {\it destabilise} the mode at low viscosity, they {\it stabilise} it at higher viscosities. On the other hand, we find the poloidal flow to be always stabilising in contrast to the no-flow case.

\begin{figure}[!htb]
\centering
\includegraphics[scale=0.35]{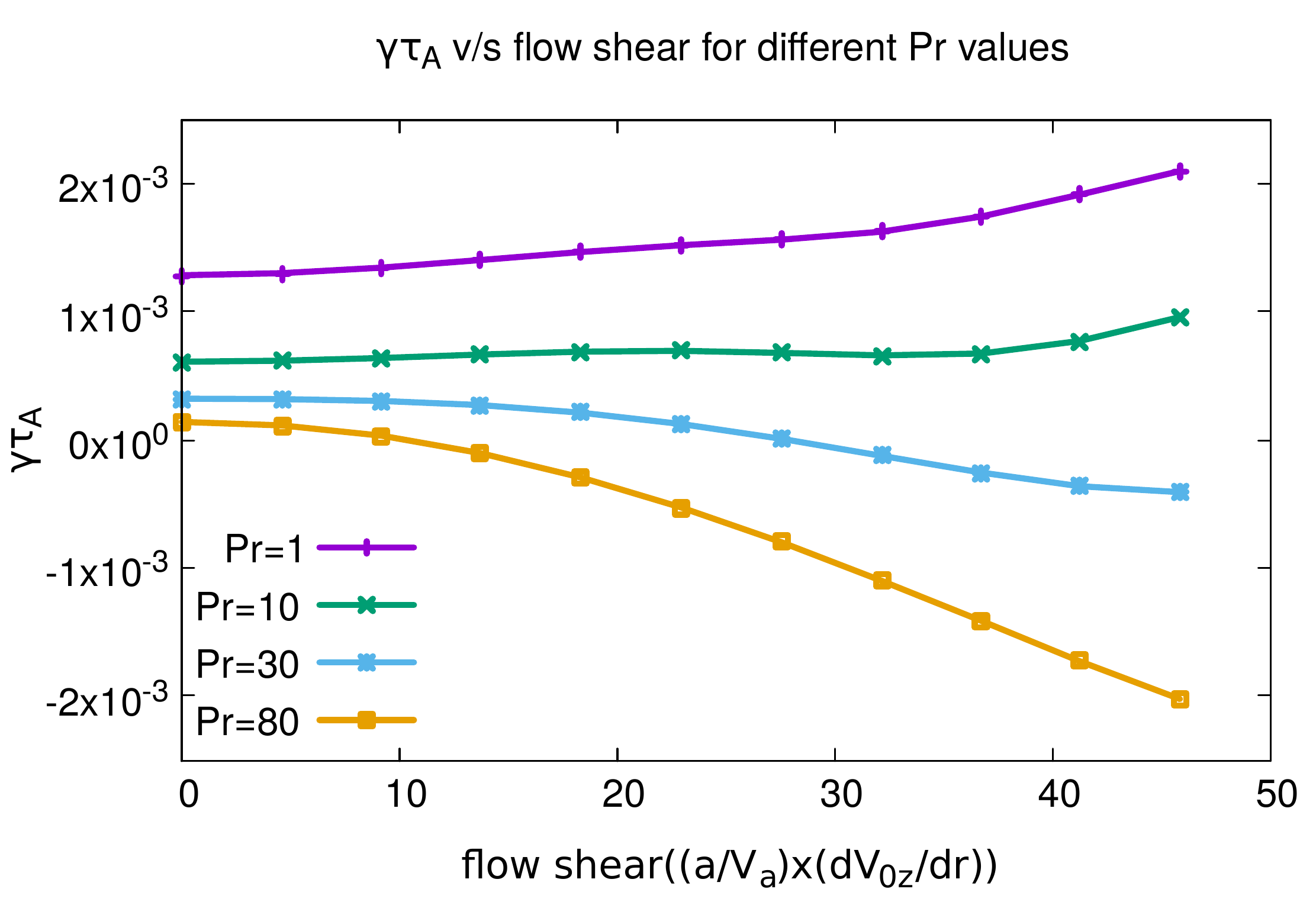} 
\caption{Linear growth rate of (1,1) mode v/s axial flow shear for different Pr values and $S=10^6$}
\label{axial_flow_diff_Pr}
\end{figure}

In Fig. \ref{axial_flow_diff_Pr},  we have shown how the linear growth rate of the $m=1, n=1$ mode changes as we go from low to high axial flow shear, $\frac{a}{V_{a}}\frac{dv_{0z}}{dr}$, for different viscosity regimes. We see that for a fixed Pr, the nature of stabilisation of flow does not change with the amount of flow shear. However, the nature of stabilisation changes depending on the viscosity regime irrespective of the amount of flow shear.

\section{Nonlinear Results}\label{nonlinear results}

Next we report on our nonlinear results for the (1,1) mode in the absence and presence of flows.  We have continued with the same parameters and related profiles that we have used for the linear runs. Here, we have a slight difference in the method of calculation as compared to the linear case. We set up an equilibrium from the given initial parameters and in every iteration we solve both the mean, i.e., (0,0) Fourier components and the perturbed components. As a result, the equilibrium evolves with time, while in the linear case, the equilibrium was held fixed.

\subsection{Axial flow}

In this section, we describe the nonlinear evolution of (1,1) modes in the presence of axial flows. Here, two different flow profiles are employed to elucidate the dependence of the results on the profile. At first, to understand the effect of flow shear we have used a tanh flow profile. The form of the tanh profile has been described in section \ref{linear axial}, but here we have used  $M_{z}=0.01$

\begin{center}
\begin{figure}[!htb]
\centering
\includegraphics[scale=0.35]{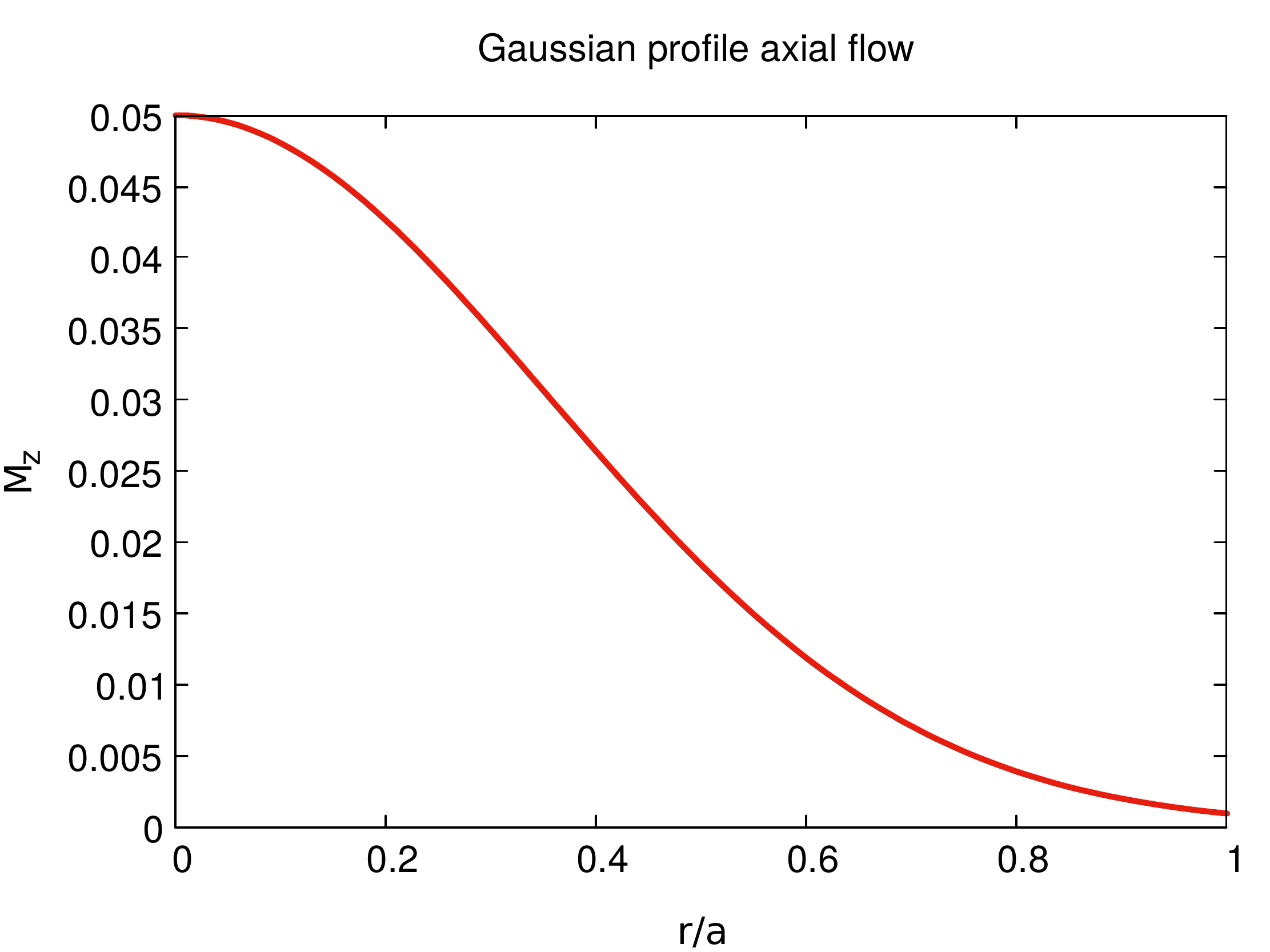} 
\caption{Axial flow profile(Gaussian profile)}
\label{axial_profile_gaussian}
\end{figure}
\end{center}

Next, we use a Gaussian flow profile which is more realistic from an experimental point of view. This profile has the form (illustrated in Fig. \ref{axial_profile_gaussian}) :

\begin{equation}
V_{0z}/V_{A}=M_{z}e^{-\rho^{2}}
\end{equation}

where, $\rho_{res}$ is the location of the mode resonant surface and $M_{z}=0.05$. 

\begin{center}
\begin{figure}[!htb]
\centering
\includegraphics[scale=0.33]{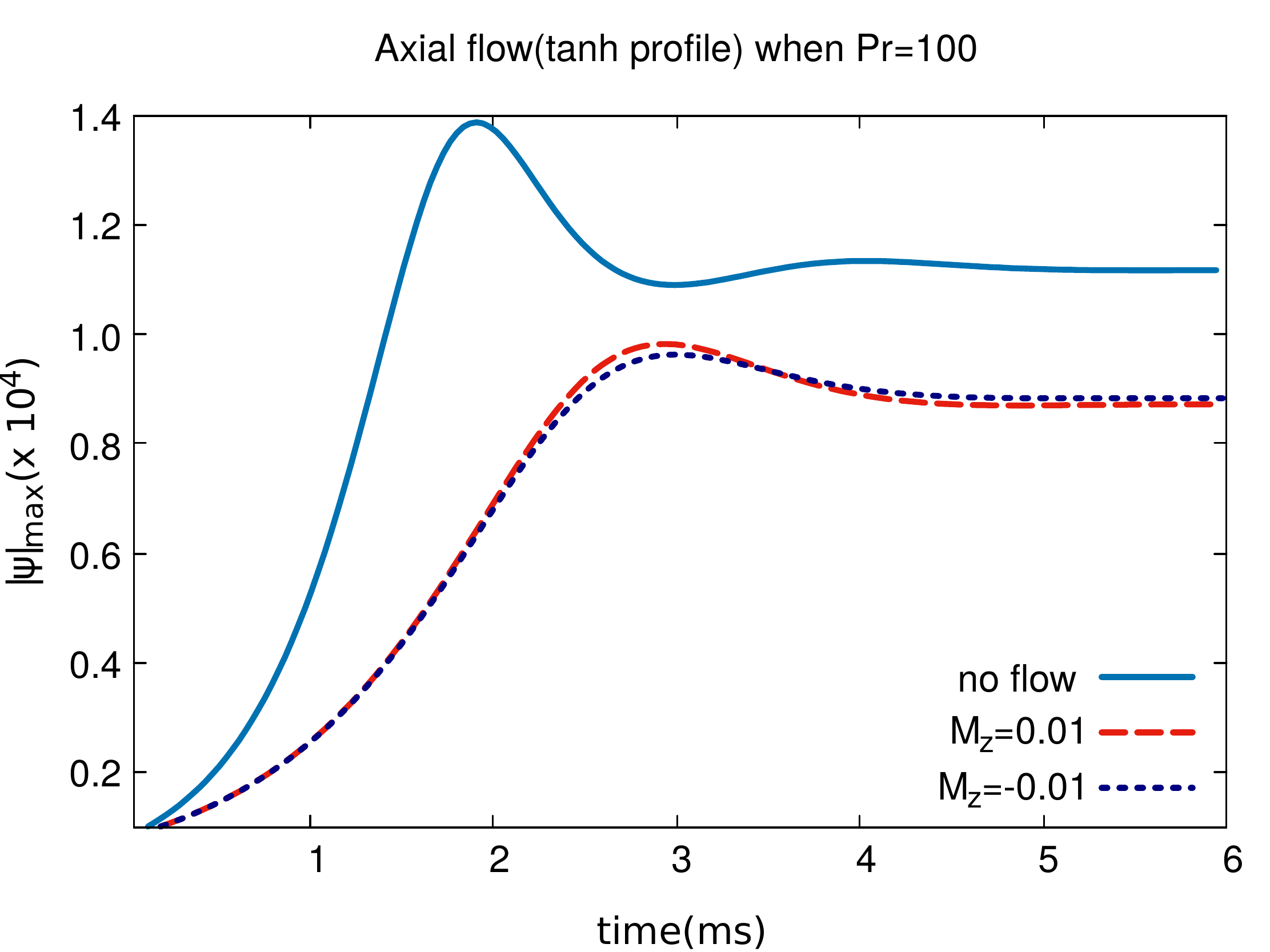}  
\caption{$\tilde{|\psi|}_{max}$ evolution with axial flow with tanh profile, $M_{z}=0.01$, $Pr=100$ and $S=10^6$.}
\label{fig:tanh_nl_12}
\end{figure}
\end{center}

\begin{center}
\begin{figure}[!htb]
\centering
\includegraphics[scale=0.33]{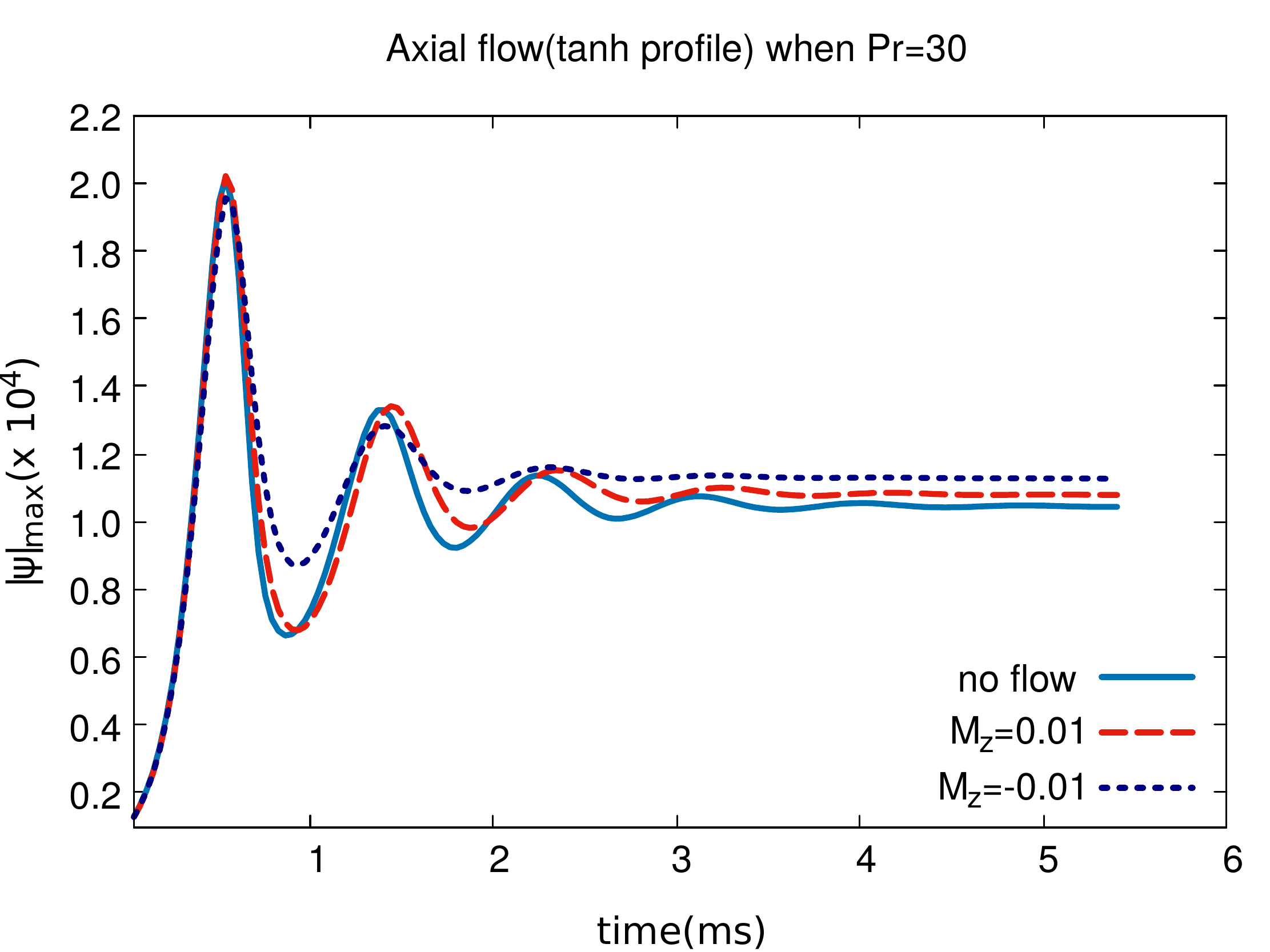}
\caption{$\tilde{|\psi|}_{max}$ evolution with axial flow with tanh profile, $M_{z}=0.01$, $Pr=30$ and $S=10^6$.}
\label{tanh_nl_pr30}
\end{figure}
\end{center}

The Figs. \ref{fig:tanh_nl_12} and \ref{tanh_nl_pr30} illustrate the time evolution of $\tilde{|\psi|}_{max}$ with a tanh flow profile for $Pr=100$ and $Pr=30$ respectively. For the high viscosity case, we notice a strong stabilisation of the (1,1) mode in the presence of axial flow both in the linear growth rate as well as in the nonlinear saturation level. However, for the low viscosity case, there is a slight increase of nonlinear saturation level of the modes in the presence of axial flow. Similar to the linear runs, the nonlinear evolution runs also show the destabilising trend of the mode for lower viscosity and a stabilising influence for higher viscosity compared to the no flow case.

Figs. \ref{gau} and \ref{fig:gauss_nl_pr30} show the nonlinear evolution of the mode with a Gaussian flow profile for $Pr=100$ and $Pr=30$ respectively. In this case, the nature of the effects is qualitatively similar to that of the tanh flow case. However, the changes in the growth rates compared to the no flow case are smaller even if the amount of flow is higher in this case. We have seen that even if the linear evolution does not depend on the sign of the flow, the nonlinear saturation levels of $\tilde{|\psi|}_{max}$ are different for different signs of flows except for the $Pr=100$ tanh flow profile case(cf. Fig. \ref{fig:tanh_nl_12}), where the difference is very small.

\begin{center}
\begin{figure}[!htb]
\centering
\includegraphics[scale=0.33]{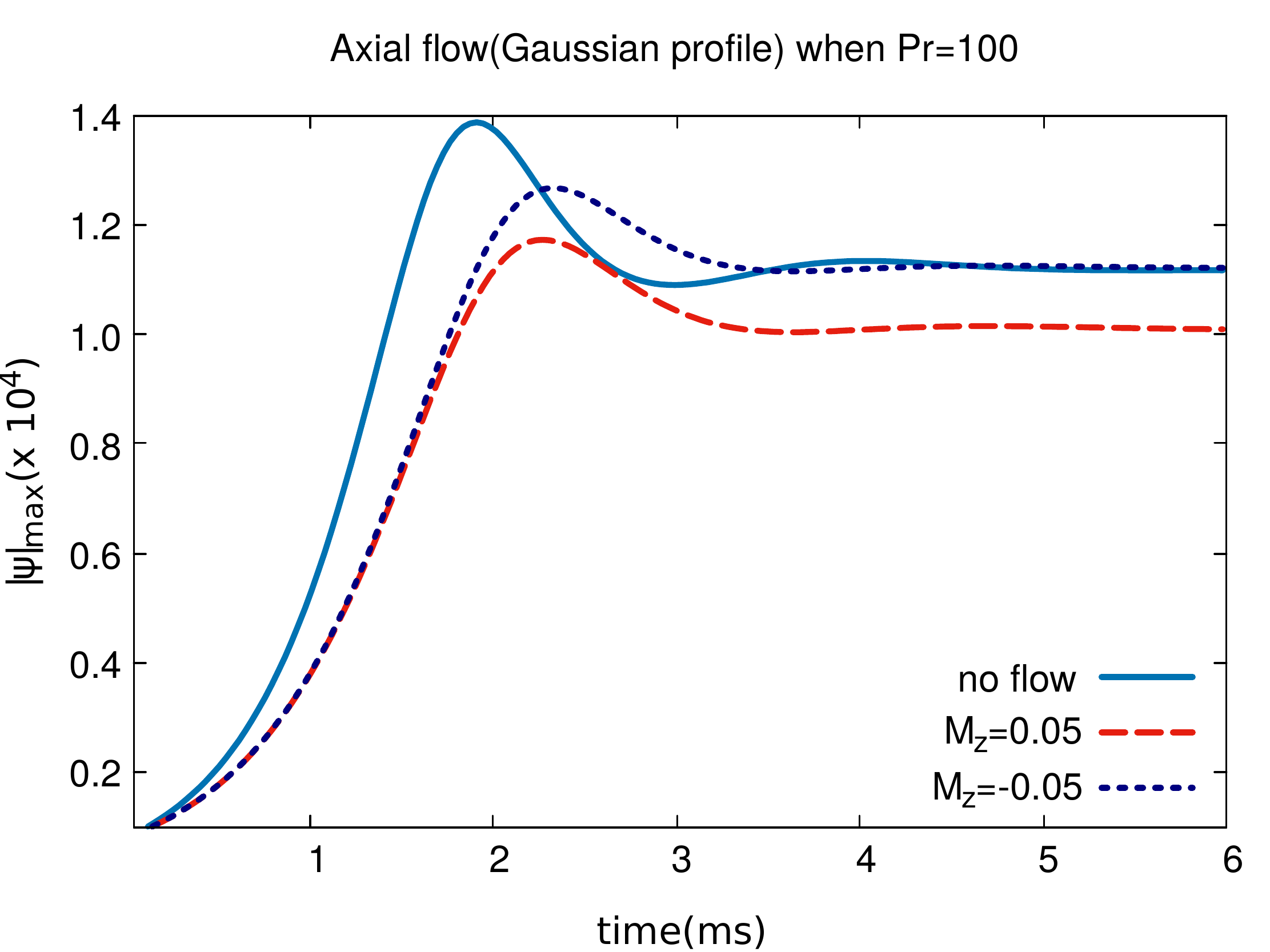} 
\caption{$\tilde{|\psi|}_{max}$ evolution with axial flow with gaussian profile, $M_{z}=0.05$, $Pr=100$ and $S=10^6$.}
\label{gau}
\end{figure}

\begin{figure}
\centering
\includegraphics[scale=0.33]{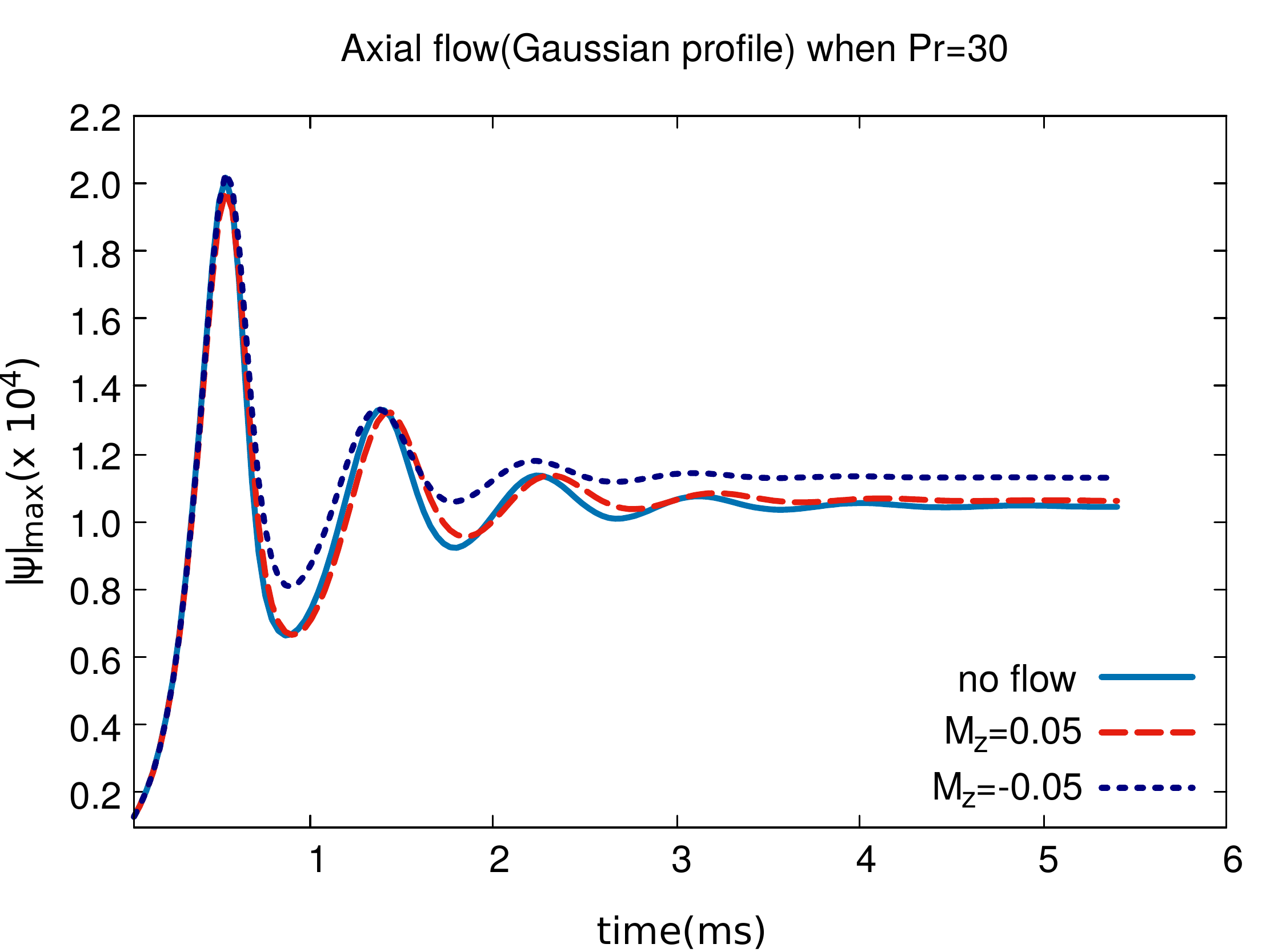}  
\caption{$\tilde{|\psi|}_{max}$ evolution with axial flow with gaussian profile, $M_{z}=0.05$, $Pr=30$ and $S=10^6$}
\label{fig:gauss_nl_pr30}
\end{figure}
\end{center}

\subsection{Poloidal Flow}

\begin{center}
\begin{figure}[!htb]
\centering
\includegraphics[scale=0.33]{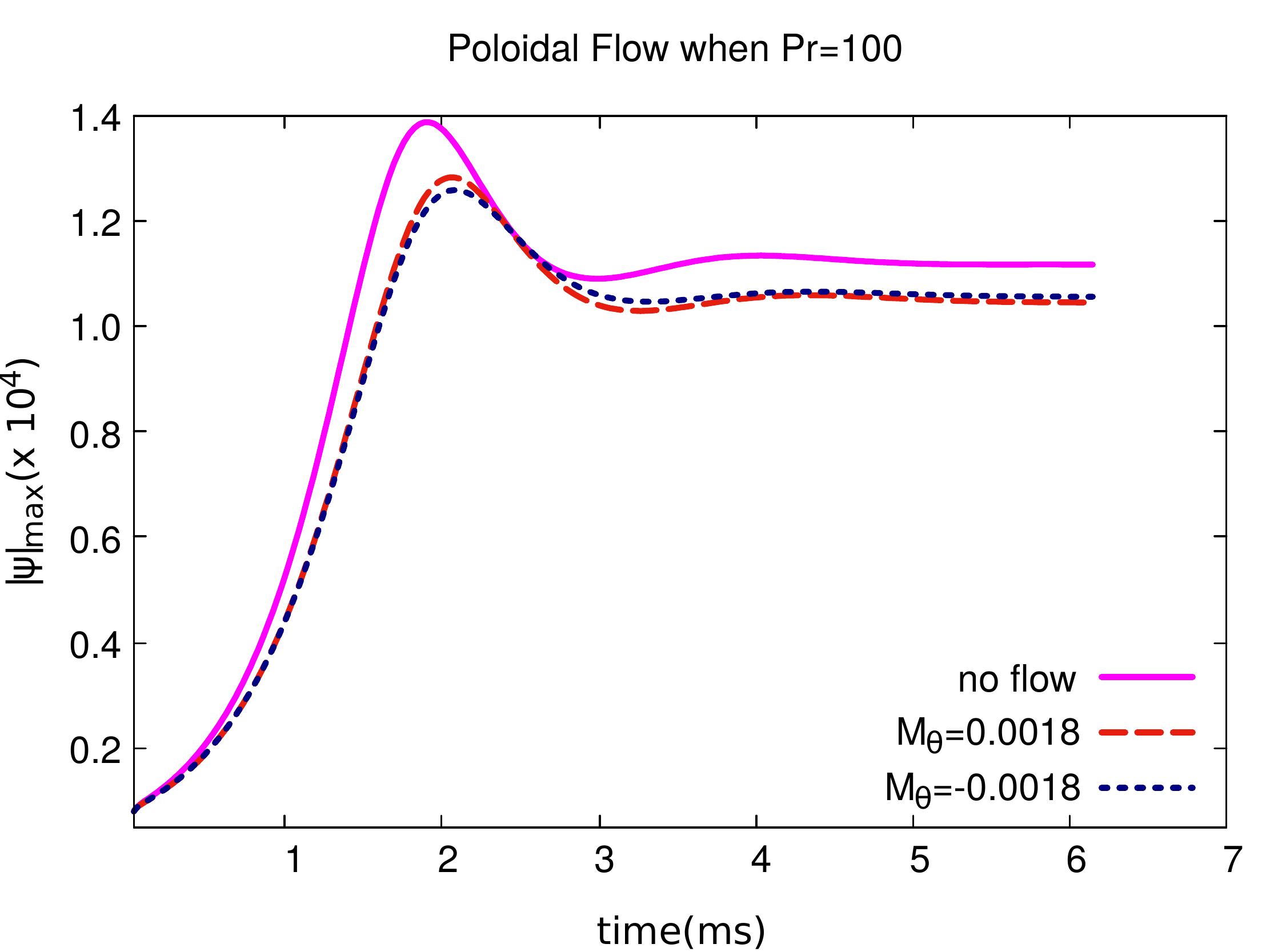}  
\caption{$\tilde{|\psi|}_{max}$ evolution with poloidal flow, $M_{\theta}=0.0018$ $Pr=100$ and $S=10^6$.}
\label{pol_flow_nl}
\end{figure}
\end{center}

In Fig. \ref{pol_flow_nl}, we display the effects of poloidal flow upon the nonlinear evolution of the amplitude of the (1,1) mode. Here, we notice that the poloidal flow stabilises the mode, and the final saturation levels are nearly equal for such small amounts of flow. For higher values of $M_{\theta}$ the saturation levels do differ significantly as a function of the direction of flow.

\begin{center}
\begin{figure}[!htb]
\centering
\includegraphics[scale=0.33]{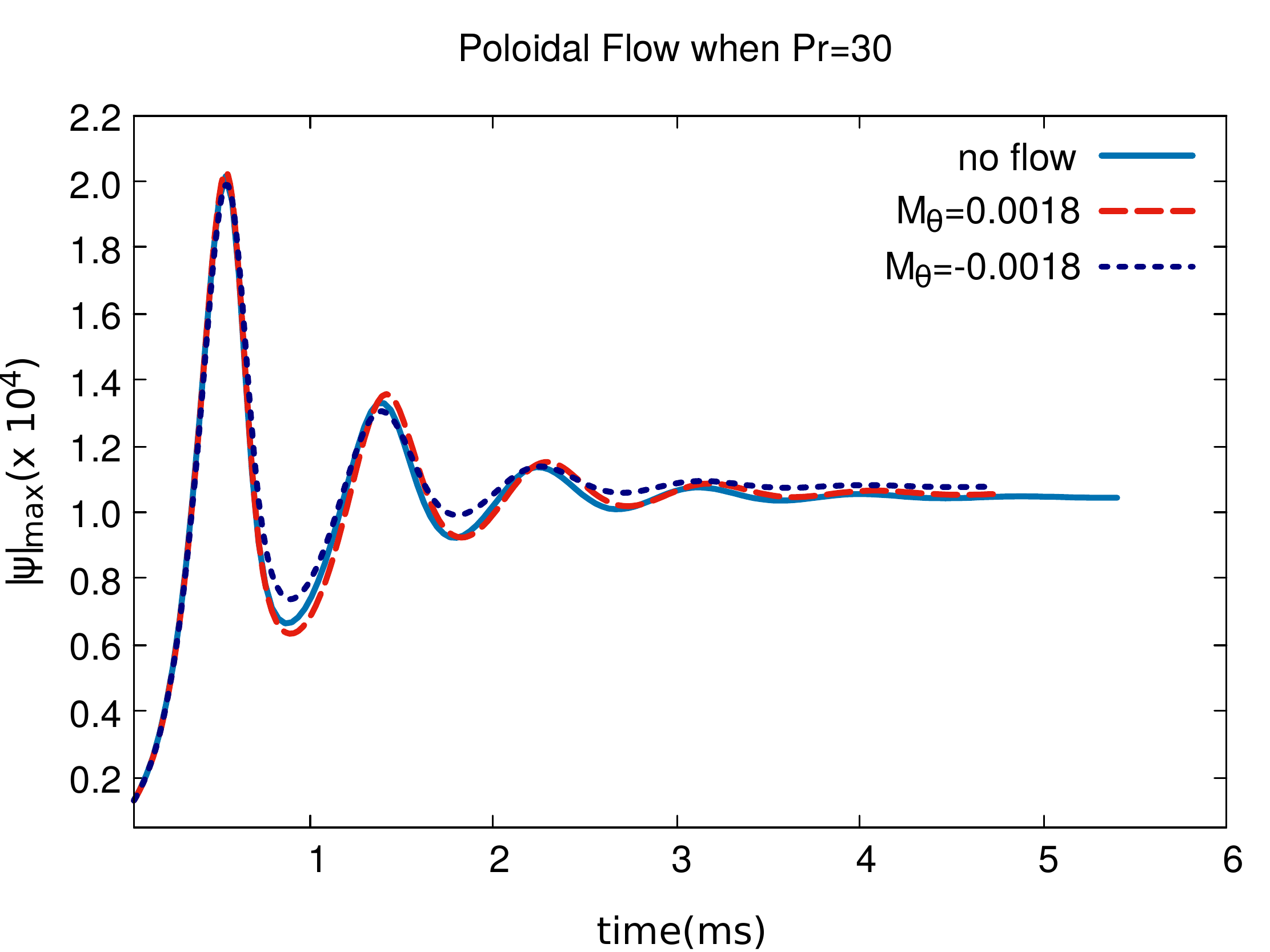} 
\caption{$\tilde{|\psi|}_{max}$ evolution with poloidal flow, $M_{\theta}=0.0018$, $Pr=30$ and $S=10^6$.}
\label{pol_flow_nl_pr30}
\end{figure}
\end{center}

We have repeated this study with a $Pr=30$ in Fig. \ref{pol_flow_nl_pr30}, and we notice here that the nonlinear saturated levels show a different behaviour from that at a higher $Pr$, in that poloidal flow now slightly destabilises the mode. The implication of this effect is clearly reflected for helical flows which we will discuss next.

\subsection{Helical Flow}

\begin{center}
\begin{figure}[!htb]
\centering
\includegraphics[scale=0.33]{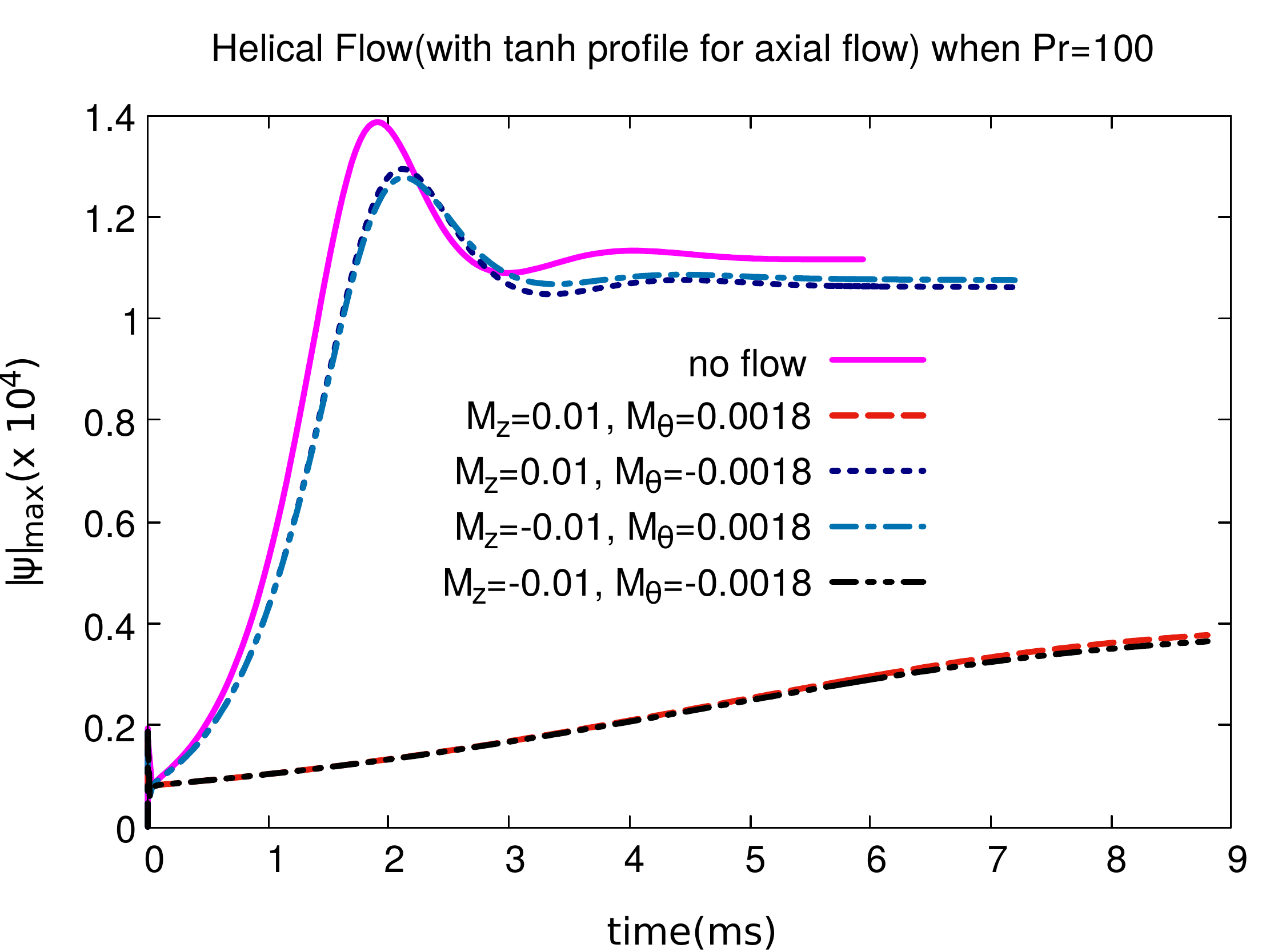}    
\caption{$\tilde{|\psi|}_{max}$ evolution with helical flow using tanh profile, $M_{z}=0.01$,   $M_{\theta}=0.0018$, $Pr=100$ and $S=10^6$.}
\label{hel_flow_tanh_nl}
\end{figure}
\end{center}

In this subsection we discuss the combined effect of axial and poloidal flows on the stability of the (1,1) mode. We have considered all four sign combinations of the axial and poloidal flows to understand the effect of flow helicity on the evolution of the mode. In Fig. \ref{hel_flow_tanh_nl}, we show the effect of a sheared axial flow with a tanh profile combined with a sheared poloidal flow for $Pr=100$. Here we can have two types of flow helicity depending on the signs of the axial flow and the poloidal flow. We find that although both the flow helicity cases impart a stabilising effect compared to the no flow case, the degree of stabilisation is very different for different flow helicities. For example, having kept the poloidal flow sign to be positive but changing the direction of the axial flow from positive to negative, both the linear growth rates and the nonlinear saturation levels have increased significantly to a much higher value. Thus, we find an asymmetry in the nature of the stabilisation of the (1,1) mode in the presence of helical flows that depends on the type of flow helicity. The change in the degree of stabilization for different helicities arises from the relationship between the flow direction and the direction of the magnetic field which essentially changes the relative sign between $q_{res}^{\prime}$ (the magnetic shear) and $v_{0,res}^{\prime}$ (the flow shear) near the mode resonant surface \cite{Haye2009}.

\begin{center}
\begin{figure}[!htb]
\centering
\includegraphics[scale=0.33]{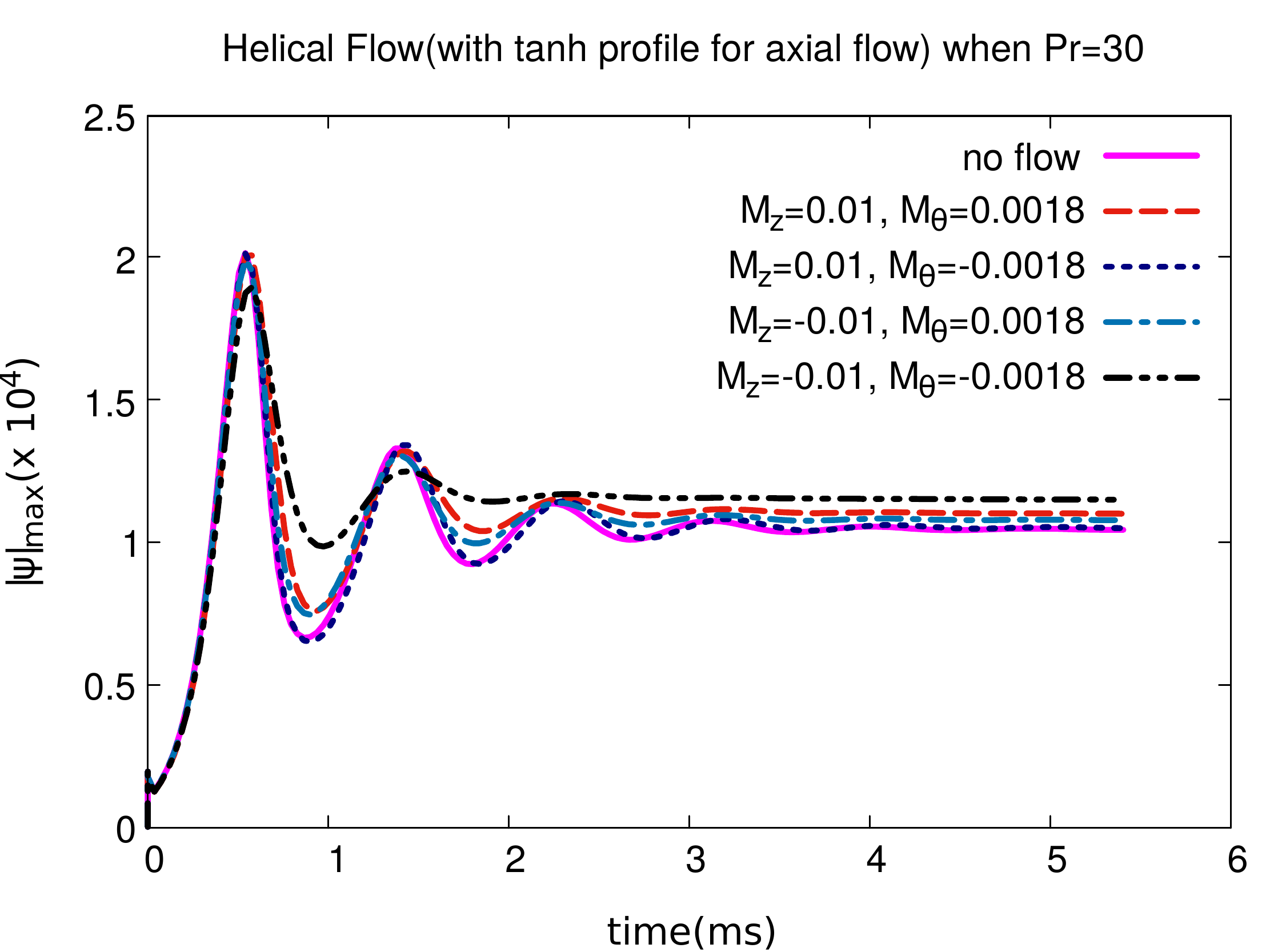}   
\caption{$\tilde{|\psi|}_{max}$ evolution with helical flow using tanh profile, $M_{z}=0.01$,     $M_{\theta}=0.0018$, $Pr=30$ and $S=10^6$.}
\label{hel_flow_tanh_nl_pr30}
\end{figure}
\end{center}

\begin{center}
\begin{figure}[!htb]
\centering
\includegraphics[scale=0.33]{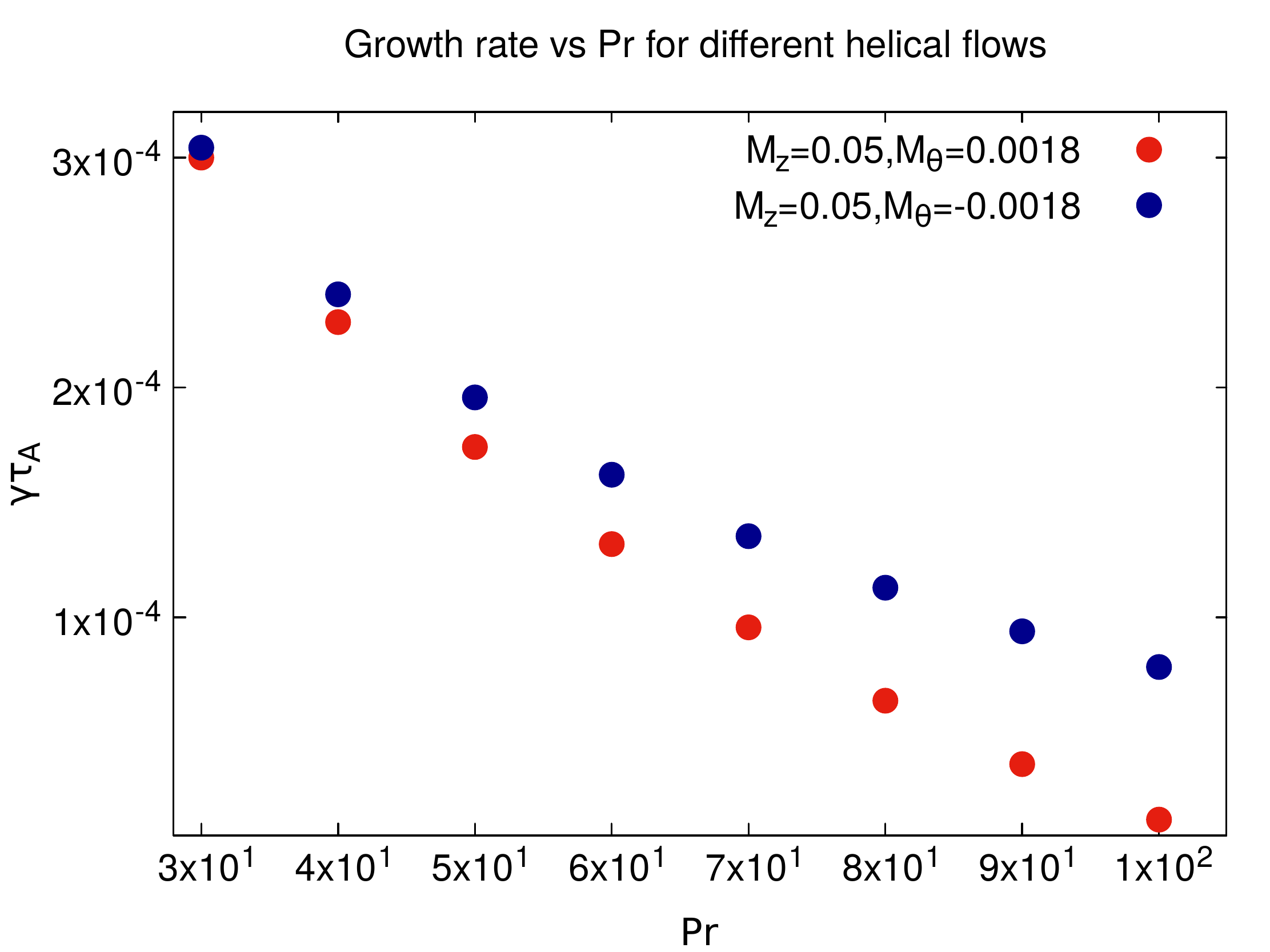} 
\caption{$\tilde{|\psi|}_{max}$ Linear growth rates with helical flows for different Prandtl numbers using tanh profile and $S=10^6$.}
\label{gr_heli_Pr}
\end{figure}
\end{center}

We have carried out a similar study at a lower viscosity of $Pr=30$ as shown in Fig. \ref{hel_flow_tanh_nl_pr30}. Here, the nonlinear saturation levels in all flow cases are slightly higher compared to the no flow case. Also, the symmetry breaking for two different flow helicities are so small both in the linear and nonlinear regime that it cannot be distinguished from the figure, but can be distinguished from numerical values of the linear growth rates and nonlinear saturation levels. In fact, the symmetry breaking effect begins to manifest itself in the linear stage itself as can be clearly seen in the difference of the slopes of the time evolution of $\mid \psi \mid_{max}$ for the two helicities of the flow. A comparison of Figs. (\ref{hel_flow_tanh_nl}) and (\ref{hel_flow_tanh_nl_pr30}) also raises the interesting question whether there occurs a ``bifurcation'' of the saturated states at some value of the Prandtl number between 30 and 100.  To check this interesting question we have numerically determined the linear growth rates for a number of different magnitudes of the helical flow and plotted their values for two different helicities in Fig.~(\ref{gr_heli_Pr}). As can be clearly seen there is a continuous transition in the behaviour as a function of $Pr$ that is indicative of an absence of any bifurcation phenomenon. 

\begin{center}
\begin{figure}[!htb]
\centering
\includegraphics[scale=0.33]{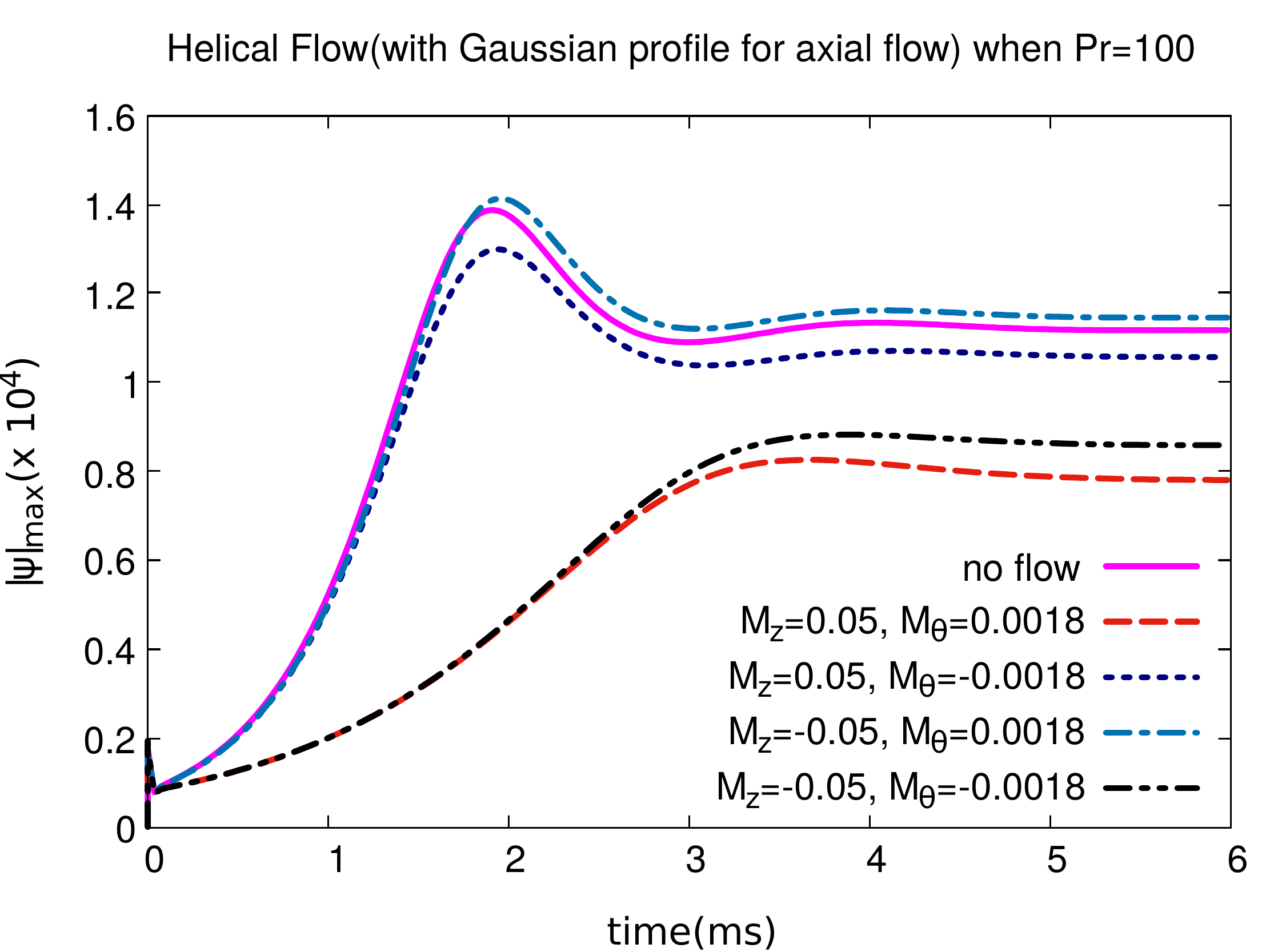}  
\caption{$\tilde{|\psi|}_{max}$ evolution with Helical flow using gaussian profile,  $M_{z}=0.05$,$M_{\theta}=0.0018$, $Pr=100$ and $S=10^6$.}
\label{hel_flow_gauss_nl}
\end{figure}
\end{center}

\begin{center}
\begin{figure}[!htb]
\centering
\includegraphics[scale=0.33]{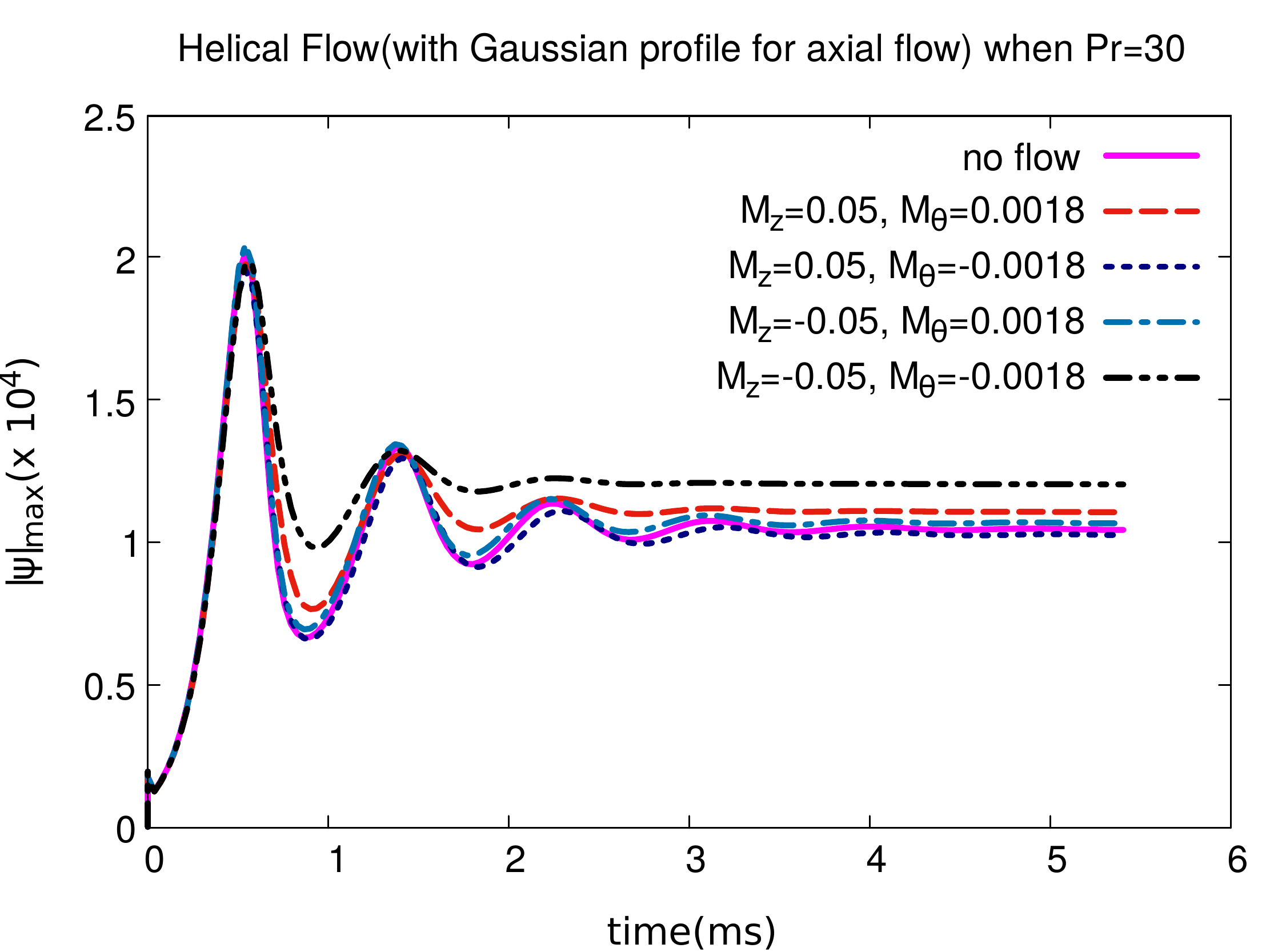}
\caption{$\tilde{|\psi|}_{max}$ evolution with helical flow using gaussian profile, $M_{z}=0.05$,$M_{\theta}=0.0018$, $Pr=30$ and $S=10^6$.}
\label{hel_flow_gauss_nl_pr30}
\end{figure}
\end{center}

In Fig. \ref{hel_flow_gauss_nl} and Fig. \ref{hel_flow_gauss_nl_pr30}, we have shown the effect of a sheared axial flow with a Gaussian profile combined with a sheared poloidal flow for $Pr=100$ and $Pr=30$ respectively. Here, we notice that the effects are very similar to the tanh flow case, but the changes in the linear growth rates and nonlinear saturation levels are relatively smaller.

On comparison of the helical flow results obtained here with those using pure axial and poloidal flows, as discussed in the previous sections, there is no symmetry breaking in the linear growth rates of the (1,1) mode if we change the direction of the flow. However, there is a difference in the nonlinear saturation levels even for those cases where we use a pure axial or poloidal flow. This is due to the self-generation of nonlinear helical terms even if we start with pure flows, as discussed in Chandra et. al. \cite{Chandra2015}.

\section{Summary and Discussion}\label{discussion}

To summarise, we have carried out linear and nonlinear studies of the (1,1) resistive internal kink mode using a V-RMHD version of the CUTIE code, in a cylindrical geometry with periodic boundary conditions. We have studied the effect of equilibrium sheared flows on the (1,1) mode and the role of viscosity in modifying the effect of flows. Viscosity can be significantly enhanced due to turbulence in tokamaks and it is expected that the $Prandtl$ $number$ can be as high as 100 \cite{Maget12, Takeda2008} in advanced scenarios for JET and ITER. %

Our results can be summarised as follows. In the linear regime our scaling studies  in the absence of flow agree with analytical results in the literature\cite{Porcelli1987}. The presence of poloidal flow does not change the linear scaling results but axial flows do bring about a significant change. We further find that the effect of viscosity on the growth rate of the mode can be significantly altered by the presence of flows. Helical flows exhibit a strong symmetry breaking with respect to the direction of the flow at high Pr but such an effect weakens at low Pr. In the nonlinear regime, for axial flows the saturation level of the mode  decreases at a higher viscosity compared to the case of no flow but slightly increases at lower viscosity. Similar results are found for the poloidal flow case. In the case of helical flows at high viscosity, there is a significant change in the nonlinear saturation level depending on the flow helicity. Such an asymmetry effect is much weaker in the low viscosity case. It might be worth mentioning that similar asymmetric effects in the sawteeth time period have been observed in tokamak experiments with a change in the direction of the equilibrium flow induced by neutral beam injections\cite{Chapman2007a, Nave2006,Chapman2006a, Chapman2008c}. Our results can prove useful in developing appropriate theoretical models for sawteeth behaviour in the presence of sheared flows and high viscosity.

\bibliographystyle{ieeetr}
\bibliography{new_rev_unmarked}

\begin{thebibliography}{10}

\bibitem{VonGoeler1974}
S.~{Von Goeler}, W.~Stodiek, and N.~Sauthoff, ``{Studies of internal
  disruptions and m=1 oscillations in tokamak discharges with soft-X-ray
  tecniques},'' {\em Physical Review Letters}, vol.~33, no.~20, pp.~1201--1203,
  1974.

\bibitem{Kadomtsev}
B.~B. {Kadomtsev}, ``{Disruptive instability in Tokamaks},'' {\em Fizika
  Plazmy}, vol.~1, pp.~710--715, Sept. 1975.

\bibitem{rosenblu}
M.~N. Rosenbluth, R.~Y. Dagazian, and P.~H. Rutherford, ``Nonlinear properties
  of the internal m = 1 kink instability in the cylindrical tokamak,'' {\em
  The Physics of Fluids}, vol.~16, no.~11, pp.~1894--1902, 1973.

\bibitem{Porcelli1996}
F.~{Porcelli}, D.~{Boucher}, and M.~N. {Rosenbluth}, ``{Model for the sawtooth
  period and amplitude},'' {\em Plasma Physics and Controlled Fusion}, vol.~38,
  pp.~2163--2186, Dec. 1996.

\bibitem{Monticello1986}
D.~A. Monticello, W.~Park, R.~Izzo, and K.~McGuire, ``{A review of calculations
  of the resistive internal m = 1 mode in Tokamaks},'' {\em Computer Physics
  Communications}, vol.~43, no.~1, pp.~57--67, 1986.

\bibitem{Rice2007}
J.~Rice, A.~Ince-Cushman, J.~deGrassie, L.-G. Eriksson, Y.~Sakamoto,
  A.~Scarabosio, A.~Bortolon, K.~Burrell, B.~Duval, C.~Fenzi-Bonizec,
  M.~Greenwald, R.~Groebner, G.~Hoang, Y.~Koide, E.~Marmar, A.~Pochelon, and
  Y.~Podpaly, ``Inter-machine comparison of intrinsic toroidal rotation in
  tokamaks,'' {\em Nuclear Fusion}, vol.~47, no.~11, p.~1618, 2007.

\bibitem{Menard2003}
J.~Menard, M.~Bell, R.~Bell, E.~Fredrickson, D.~Gates, S.~Kaye, B.~LeBlanc,
  R.~Maingi, D.~Mueller, S.~Sabbagh, D.~Stutman, C.~Bush, D.~Johnson, R.~Kaita,
  H.~Kugel, R.~Maqueda, F.~Paoletti, S.~Paul, M.~Ono, Y.-K. Peng, C.~Skinner,
  E.~Synakowski, and the NSTX Research~Team, ``Beta-limiting mhd instabilities
  in improved-performance nstx spherical torus plasmas,'' {\em Nuclear Fusion},
  vol.~43, no.~5, p.~330, 2003.

\bibitem{Menard2005}
J.~E. {Menard}, R.~E. {Bell}, E.~D. {Fredrickson}, D.~A. {Gates}, S.~M. {Kaye},
  B.~P. {LeBlanc}, R.~{Maingi}, S.~S. {Medley}, W.~{Park}, S.~A. {Sabbagh},
  A.~{Sontag}, D.~{Stutman}, K.~{Tritz}, W.~{Zhu}, and {NSTX Research Team},
  ``{Internal kink mode dynamics in high-{$\beta$} NSTX plasmas},'' {\em
  Nuclear Fusion}, vol.~45, pp.~539--556, July 2005.

\bibitem{Chapman2007a}
I.~Chapman, S.~D. Pinches, J.~P. Graves, R.~J. Akers, L.~C. Appel, R.~V. Budny,
  S.~Coda, N.~J. Conway, M.~de~Bock, L.-G. Eriksson, R.~J. Hastie, T.~C.
  Hender, G.~T.~a. Huysmans, T.~Johnson, H.~R. Koslowski,
  A.~Kr{\"{a}}mer-Flecken, M.~Lennholm, Y.~Liang, S.~Saarelma, S.~E. Sharapov,
  and I.~Voitsekhovitch, ``{The physics of sawtooth stabilization},'' {\em
  Plasma Physics and Controlled Fusion}, vol.~49, pp.~B385--B394, 2007.

\bibitem{Nave2006}
M.~F.~F. Nave, H.~R. Koslowski, S.~Coda, J.~Graves, M.~Brix, R.~Buttery,
  C.~Challis, C.~Giroud, M.~Stamp, and P.~{De Vries}, ``{Exploring a small
  sawtooth regime in Joint European Torus plasmas with counterinjected neutral
  beams},'' {\em Physics of Plasmas}, vol.~13, no.~1, pp.~1--4, 2006.

\bibitem{Chapman2006a}
I.~Chapman, T.~C. Hender, S.~Saarelma, S.~Sharapov, R.~Akers, N.~Conway, and
  MAST, ``{The effect of toroidal plasma rotation on sawteeth in MAST},'' {\em
  Nuclear Fusion}, vol.~46, pp.~1009--1016, 2006.

\bibitem{Chapman2008c}
I.~Chapman, S.~Pinches, H.~Koslowski, Y.~Liang, A.~Kr{\"{a}}mer-Flecken, and
  M.~de~Bock, ``{Sawtooth stability in neutral beam heated plasmas in
  TEXTOR},'' {\em Nuclear Fusion}, vol.~48, no.~3, p.~035004, 2008.

\bibitem{Morrison12}
X.~L. {Chen} and P.~J. {Morrison}, ``{Resistive tearing instability with
  equilibrium shear flow},'' {\em Physics of Fluids B}, vol.~2, pp.~495--507,
  Mar. 1990.

\bibitem{Guzdar1234}
R.~G. Kleva and P.~N. Guzdar, ``Stabilization of sawteeth in tokamaks with
  toroidal flows,'' {\em Physics of Plasmas}, vol.~9, no.~7, pp.~3013--3018,
  2002.

\bibitem{Shumlak1995}
U.~Shumlak and C.~Hartman, ``{Sheared Flow Stabilization of the m=1 Kink Mode
  in Z Pinches},'' {\em Physical Review Letters}, vol.~75, no.~18,
  pp.~3285--3288, 1995.

\bibitem{gatto}
R.~Gatto, P.~Terry, and C.~Hegna, ``Tearing mode stability with equilibrium
  flows in the reversed-field pinch,'' {\em Nuclear Fusion}, vol.~42, no.~5,
  p.~496, 2002.

\bibitem{naitou_kobayashi_tokuda_1999}
H.~Naitou, T.~Kobayashi, and S.~Tokuda, ``Stabilization of the kinetic internal
  kink mode by a sheared poloidal flow,'' {\em Journal of Plasma Physics},
  vol.~61, no.~4, p.~543–552, 1999.

\bibitem{Mikhailovskii2008aa}
A.~B. Mikhailovskii, J.~G. Lominadze, R.~M.~O. Galv{\~a}o, A.~P. Churikov,
  N.~N. Erokhin, V.~D. Pustovitov, S.~V. Konovalov, A.~I. Smolyakov, and V.~S.
  Tsypin, ``Ideal internal kink modes in a differentially rotating cylindrical
  plasma,'' {\em Plasma Physics Reports}, vol.~34, no.~7, pp.~538--546, 2008.

\bibitem{wahl_bond}
C.~Wahlberg and A.~Bondeson, ``Stabilization of the internal kink mode in a
  tokamak by toroidal plasma rotation,'' {\em Physics of Plasmas}, vol.~7,
  no.~3, pp.~923--930, 2000.

\bibitem{wael12}
F.~L. Waelbroeck, ``Gyroscopic stabilization of the internal kink mode,'' {\em
  Physics of Plasmas}, vol.~3, no.~3, pp.~1047--1053, 1996.

\bibitem{Brunetti2017}
D.~{Brunetti}, E.~{Lazzaro}, and S.~{Nowak}, ``{Ideal and resistive
  magnetohydrodynamic instabilities in cylindrical geometry with a sheared flow
  along the axis},'' {\em Plasma Physics and Controlled Fusion}, vol.~59,
  p.~055012, May 2017.

\bibitem{Maget12}
P.~Maget, ``{Non linear MHD Modelling of NTMs in JET Advanced Scenarios},''
  {\em IAEA CONFERENCE}, pp.~3--10, 2010.

\bibitem{Wang2015}
S.~Wang and Z.~W. Ma, ``Influence of toroidal rotation on resistive tearing
  modes in tokamaks,'' {\em Physics of Plasmas}, vol.~22, no.~12, p.~122504,
  2015.

\bibitem{Tala2011}
T.~Tala, ``{Parametric dependences of momentum pinch and Prandtl number in
  JET},'' {\em Nuclear Fusion}, vol.~51, 2011.

\bibitem{Takeda2008}
K.~Takeda, O.~Agullo, S.~Benkadda, A.~Sen, N.~Bian, and X.~Garbet, ``{Nonlinear
  viscoresistive dynamics of the m=1 tearing instability},'' {\em Physics of
  Plasmas}, vol.~15, p.~022502, 2008.

\bibitem{Chen1990a}
X.~L. Chen and P.~J. Morrison, ``The effect of viscosity on the resistive
  tearing mode with the presence of shear flow,'' {\em Physics of Fluids B:
  Plasma Physics}, vol.~2, no.~11, pp.~2575--2580, 1990.

\bibitem{Ofman1991}
L.~Ofman, X.~L. Chen, P.~J. Morrison, and R.~S. Steinolfson, ``Resistive
  tearing mode instability with shear flow and viscosity,'' {\em Physics of
  Fluids B: Plasma Physics}, vol.~3, no.~6, pp.~1364--1373, 1991.

\bibitem{Ren1999}
C.~Ren, M.~S. Chu, and J.~D. Callen, ``Magnetic island deformation due to
  sheared flow and viscosity,'' {\em Physics of Plasmas}, vol.~6, no.~4,
  pp.~1203--1207, 1999.

\bibitem{Haye2009}
R.~J.~L. Haye and R.~J. Buttery, ``The stabilizing effect of flow shear on
  m/n=3/2 magnetic island width in {DIII-D},'' {\em Physics of Plasmas},
  vol.~16, no.~2, p.~022107, 2009.

\bibitem{Thyagaraja2000}
A.~Thyagaraja, ``{Numerical simulations of tokamak plasma turbulence and
  internal transport barriers},'' {\em Plasma Phys. Control. Control. Fusion},
  vol.~42, no.~00, pp.~255--269, 2000.

\bibitem{Porcelli1987}
F.~Porcelli, ``{Viscous resistive magnetic reconnection},'' {\em Physics of
  Fluids}, vol.~30, no.~6, p.~1734, 1987.

\bibitem{Chandra2015}
D.~Chandra, A.~Thyagaraja, A.~Sen, C.~Ham, T.~Hender, R.~Hastie, J.~Connor,
  P.~Kaw, and J.~Mendonca, ``Modelling and analytic studies of sheared flow
  effects on tearing modes,'' {\em Nuclear Fusion}, vol.~55, no.~5, p.~053016,
  2015.

\bibitem{Thyagaraja2010}
A.~Thyagaraja, M.~Valovi\v{c}, and P.~J. Knight, ``Global two-fluid turbulence
  simulations of l-h transitions and edge localized mode dynamics in the
  compass-d tokamak,'' {\em Physics of Plasmas}, vol.~17, no.~4, p.~042507,
  2010.

\bibitem{Chandra2017}
D.~Chandra, A.~Thyagaraja, A.~Sen, and P.~Kaw, ``Nonlinear simulation of elm
  dynamics in the presence of resonant magnetic perturbations,'' {\em Nuclear
  Fusion}, vol.~57, no.~7, p.~076001, 2017.

\bibitem{Militello2004a}
F.~Militello, G.~Huysmans, M.~Ottaviani, and F.~Porcelli, ``{Effects of local
  features of the equilibrium current density profile on linear tearing
  modes},'' {\em Physics of Plasmas}, vol.~11, pp.~125--128, 2004.

\bibitem{Gimblett1996}
C.~G. Gimblett, R.~J. Hastie, R.~A. M.~V. der Linden, and J.~A. Wesson,
  ``Non‐uniform rotation and the resistive wall mode,'' {\em Physics of
  Plasmas}, vol.~3, no.~10, pp.~3619--3627, 1996.

\end{thebibliography}

\end{document}